\documentclass[11pt, a4paper]{article}
\pdfoutput=1

\usepackage[utf8]{inputenc}
\usepackage{graphicx}
\usepackage{geometry}
\usepackage[english]{babel}

\usepackage{tikz}
\usetikzlibrary{decorations.pathmorphing}
\usepackage{pgfplots}
\pgfplotsset{compat=1.14}
\usepgfplotslibrary{patchplots}
\usepackage{subcaption} 

\usepackage[numbers,sort&compress]{natbib}
\usepackage{epsfig}
\usepackage{amsmath}
\usepackage{amssymb}
\usepackage{amsfonts}
\usepackage{anysize}
\usepackage{latexsym}
\usepackage{xcolor}
\usepackage[pdftex,colorlinks=true,linkcolor=blue,citecolor=blue]{hyperref}
\hypersetup{linktocpage}
\usepackage{float}
\usepackage{dsfont}
\numberwithin{equation}{section}

\definecolor{ccqqqq}{rgb}{1,0.5,0}
\definecolor{uuuuuu}{rgb}{0.26666666666666666,0.26666666666666666,0.26666666666666666}
\definecolor{qqwwzz}{rgb}{0,0.3,0.9}

\newcommand{\beq}{\begin{equation}}
\newcommand{\eeq}{\end{equation}}
\newcommand{\bea}{\begin{eqnarray}}
\newcommand{\eea}{\end{eqnarray}}
\newcommand{\bit}{\begin{itemize}}
\newcommand{\eit}{\end{itemize}}

\def\a{\alpha}

\def\l{\lambda}

\def\p{\partial}

\def\le{\left(}
\def\ri{\right)}

\newcommand{\sn}[2]{\text{sn}\left(\left.\! #1\right| #2\right)}

\newcommand{\cn}[2]{\text{cn}\left(\left.\! #1\right| #2\right)}

\newcommand{\dn}[2]{\text{dn}\left(\left.\! #1\right| #2\right)}

\renewcommand{\a}{\alpha}

\renewcommand{\d}{\delta}
\renewcommand{\th}{\theta}

\renewcommand{\l}{\lambda}

\newcommand{\vf}{\varphi}
\newcommand{\F}{\Phi}

\renewcommand{\t}{\tau}

\newcommand{\elF}[2]{\mathds{F}\left(\left.#1\right|#2\right)}
\newcommand{\elE}[2]{\mathds{E}\left(\left.#1\right|#2\right)}
\newcommand{\eE}{\mathds{E}}
\newcommand{\eK}{\mathds{K}}

\newcommand{\eP}[2]{\Pi\left(\left.#1\right|#2\right)}
\date{}

\begin{document}

\begin{titlepage}

\begin{flushright}

\end{flushright}
\bigskip
\begin{center}
{\LARGE  {\bf
Volume complexity  for Janus AdS$_3$ geometries
  \\[2mm] } }
\end{center}
\bigskip
\begin{center}
{\large \bf  Roberto  Auzzi$^{a,b}$},
 {\large \bf Stefano Baiguera$^{c}$},
  {\large \bf Sara Bonansea$^{c}$}, \\
  {\large \bf Giuseppe Nardelli$^{a,d}$}  {\large \bf and } 
    {\large \bf Kristian Toccacelo$^{c}$}
\vskip 0.20cm
\end{center}
\vskip 0.20cm 
\begin{center}
$^a${ \it \small  Dipartimento di Matematica e Fisica,  Universit\`a Cattolica
del Sacro Cuore, \\
Via Musei 41, 25121 Brescia, Italy}
\\ \vskip 0.20cm 
$^b${ \it \small{INFN Sezione di Perugia,  Via A. Pascoli, 06123 Perugia, Italy}}
\\ \vskip 0.20cm 
$^c${ \it \small{The Niels Bohr Institute, University of Copenhagen, \\
Blegdamsvej 17, DK-2100 Copenhagen \O, Denmark}}
\\ \vskip 0.20cm 
$^d${ \it \small{TIFPA - INFN, c/o Dipartimento di Fisica, Universit\`a di Trento, \\ 38123 Povo (TN), Italy} }
\\ \vskip 0.20cm 
E-mails: roberto.auzzi@unicatt.it, stefano.baiguera@nbi.ku.dk,  \\
sara.bonansea@nbi.ku.dk, giuseppe.nardelli@unicatt.it, \\
btm438@alumni.ku.dk
\end{center}
\vspace{3mm}

\begin{abstract}

We investigate the complexity=volume proposal in the case of
Janus AdS$_3$ geometries, both at zero and finite temperature.
The leading contribution coming from the Janus interface is a 
logarithmic divergence,  whose coefficient is a function of the dilaton excursion.  
In the presence of the defect, complexity is no longer topological and becomes temperature-dependent.
We also study the time evolution of the extremal volume for the 
time-dependent Janus BTZ black hole.
This background is not dual to an interface but to a pair of entangled CFTs with different values of the couplings. 
At late times, when the equilibrium is restored, the couplings of the CFTs do not influence the complexity rate. 
On the contrary, the complexity rate for the out-of-equilibrium system is always smaller compared to the pure BTZ black hole background.

\end{abstract}

\end{titlepage}

\tableofcontents

\section{Introduction}
\label{sect-intro}

Since the discovery of the Bekenstein-Hawking formula for the entropy, there has been a greater understanding of the quantum information properties encoded by the geometry of spacetime via holography.
One of the biggest achievements of this idea in the context of AdS/CFT was  the Hubeny-Ryu-Takayanagi (HRT) formula \cite{Ryu:2006bv,Hubeny:2007xt}, 
which relates the entanglement entropy for a subregion on the boundary
 to the area of a codimension-two extremal surface $\gamma$ anchored  to the boundary of the subsystem, that is,
\beq
S_A = \frac{A(\gamma)}{4G} \, . 
\eeq
This holographic proposal was justified from the perspective of the gravitational path integral \cite{Lewkowycz:2013nqa} and later generalized to include quantum corrections \cite{Faulkner:2013ana}.
These advancements led to the concept of generalized extremal surfaces, a prescription where the fine-grained entropy 
of a system is computed by extremizing the sum of the HRT entropy and a contribution which accounts for the quantum corrections due to matter fields.
In particular, this conjecture was used to
 recover the Page curve for evaporating black holes \cite{Almheiri:2019psf, Penington:2019npb, Almheiri:2019hni}.

The area of codimension-two extremal surfaces plays a central role in holography.
It is then natural to wonder about the meaning of  the volume of codimension-one objects,
whose role is not yet well established in holography.
The most promising proposal 
 is, arguably, that of holographic complexity \cite{Susskind:2014rva}. Other related possibilities include the concepts of
 quantum information metric and fidelity
\cite{MIyaji:2015mia, Alishahiha:2015rta}.

Motivated by the desire of finding a dual description to the growth of the Einstein-Rosen Bridge (ERB)
in black holes,  Susskind and collaborators \cite{Susskind:2014rva,Stanford:2014jda,Susskind:2014moa} 
introduced the notion of computational complexity in holography.
By studying the time evolution of wormholes, e.g., an eternal Black Hole (BH) 
in asymptotically AdS spacetime, one can figure out that spacelike slices connecting the two asymptotic
 boundaries evolve for a much longer time compared to the thermalization time scale.
Since these slices can be defined in a coordinate-invariant way by selecting the one with maximal volume at each value of time, 
 it is expected that there exists a dual quantity on the field theory side.
In \cite{Susskind:2014rva} it was conjectured that the dual is computational complexity, 
which is heuristically defined as a measure of the minimal number of simple unitary operations relating 
a reference state with a chosen target state.
For quantum mechanical systems, complexity can be defined geometrically as a Hamiltonian control problem, an approach pioneered by Nielsen et al. \cite{Nielsen1,Nielsen2}; in this approach, negative curvatures seem to be a desirable feature of complexity space \cite{Brown:2016wib,Brown:2019whu,Auzzi:2020idm}.
While there has been some recent progress in extending
Nielsen's method to free field theory \cite{Jefferson:2017sdb,Chapman:2017rqy, Khan:2018rzm},
a precise definition of complexity in interacting CFTs is still lacking.
Further developments in this direction can be found in, e.g., \cite{Caputa:2018kdj,Erdmenger:2020sup,Chagnet:2021uvi}.
Other approaches have also gained  attention in the community, see for instance
 \cite{Caputa:2017yrh, 
Parker:2018yvk,
 Barbon:2019wsy}.

Two proposals have made their way as holographic duals to computational complexity:
\begin{itemize}
\item the Complexity=Volume conjecture (CV), where the complexity is proportional to the volume of a maximal 
spacelike codimension-one surface $\Gamma$ anchored at the boundary
\beq
\mathcal{C}_V \sim \frac{V(\Gamma)}{G L} \, ,
\eeq
where $L$ is the AdS radius. 
This proposal has been tested in geometries with shock waves perturbations and reproduced
 phenomena expected from tensor networks and quantum circuits \cite{Stanford:2014jda}; and 
\item the Complexity=Action conjecture (CA).
Since the CV conjecture is not universal (it contains a specific dependence from a length scale like the AdS radius), it was later proposed that the complexity is holographically given by the gravitational action $I$ computed in the Wheeler-De Witt (WDW) patch, i.e., the bulk domain of dependence of the ERB \cite{Brown:2015bva, Brown:2015lvg}:
\beq
\mathcal{C}_A = \frac{I_{\mathrm{WdW}}}{\pi} \, .
\eeq
The late time behaviour of volume and action for asymptotically AdS black holes is found to be 
qualitatively the same, but they differ for intermediate times \cite{Carmi:2017jqz}.
\end{itemize}

The previous proposals have natural generalisation to the case of mixed states.
In the CV case, the proposal is to compute the volume of a maximal slice anchored to a boundary subregion and bounded by the corresponding HRT surface \cite{Alishahiha:2015rta}: hence the name \textit{subregion complexity}.
 To compute subregion CA,   it was proposed \cite{Carmi:2016wjl}
to evaluate the gravitational  action on the intersection of the WDW patch and the entanglement wedge.  
Subregion complexity for an arbitrary interval in a BTZ black hole  \cite{Banados:1992gq} background
 is an example where volume and action proposals behave differently \cite{Abt:2017pmf, Auzzi:2019vyh}.
 Further studies on subregion complexity include \cite{Abt:2018ywl,Agon:2018zso,Auzzi:2019fnp,Chen:2018mcc,Auzzi:2019mah,Caceres:2019pgf}.

Defects, interfaces and boundaries are often present in interesting physical systems.
Impurities are one of the main topics in condensed matter physics.
From a field theory perspective, we can model these systems
as a   renormalization group (RG) flow where bulk 
and boundary degrees of freedom are coupled. Entanglement entropy also provides an important tool to characterize the properties of
defects under RG flow. The entanglement entropy $S$
of an interval with length $l$ in a two-dimensional CFT centered around the defect
has the following form \cite{Calabrese:2004eu}
\beq
S=\frac{c}{3} \log \frac{l}{\delta} +\log g \, ,
\eeq
where  $c$ is the central charge, $\delta$ the UV cutoff, and $g$
the ground state degeneracy of the defect.
The quantity $g$ is monotonic under RG flow \cite{Affleck:1991tk}
and gives a measure of the number of degrees of freedom localised on the defect.

Holographic examples of geometries dual to defect CFTs include:
\begin{itemize}
\item the  Randall-Sundrum model (RS)
\cite{Randall:1999vf}. The simplest situation corresponds to a two-dimensional brane
embedded in AdS$_3$, as considered in \cite{Azeyanagi:2007qj} for the study of entanglement
entropy. The case of a conformal defect in $d$ dimensions was recently investigated in the context of the island conjecture in relation to subregion complexity \cite{Bhattacharya:2021jrn};
\item the holographic dual of CFT with boundaries (BCFT) studied in
 \cite{Takayanagi:2011zk,Fujita:2011fp, Flory:2017ftd}; 
 and
 \item  Janus geometries \cite{Bak:2003jk,Bak:2007jm}.
While it was originally discovered as a non-supersymmetric dilatonic deformation of $\mathrm{AdS}_5$ space \cite{Bak:2003jk},
 it was later found that a similar deformation can be implemented on
   $\mathrm{AdS}_3 \times S^3 \times M_4 $, where $M_4$ is a compact manifold \cite{Bak:2007jm}.
Since these solutions can be embedded into type IIB supergravity, they  have well-defined top-down holographic duals.
 \end{itemize}
All these  geometries are interesting laboratories to investigate
quantum information aspects of the AdS/CFT duality.
The Janus geometry was recently considered as a background to test the pseudo-entropy,
 a quantum information object which is conjectured to describe a Euclidean version 
 of the entanglement entropy \cite{Nakata:2020fjg}.

The physics of defects may provide us deeper insights on holographic complexity.
While this was  already studied previously for the
three-dimensional RS model  in \cite{Chapman:2018bqj}
and for BCFT  \cite{Sato:2019kik, Braccia:2019xxi},
the purpose of this paper is to fill the gap
for three-dimensional Janus geometries. Related studies focusing on fidelity can be found in
\cite{MIyaji:2015mia, Bak:2015jxd, Mazhari:2016yng}. 
These backgrounds represent an important playground to test the behavior of holographic complexity because
they admit many deformations with well-defined holographic duals, including the finite temperature case
or time-dependent perturbations.

In section \ref{sect:janus_review} we review some basic facts
about Janus geometries in three dimensions.
In section \ref{sect-CV_Janus} we compute the 
subregion volume complexity for  a symmetric region with radius $l/2$ around the defect
in  Janus AdS$_3$. The pure state complexity
can be extracted from the $l \rightarrow \infty$ limit
of this result, where the length $l$ plays the role of an IR regulator.
In section \ref{section-static-BTZ} we  extend the calculation at finite temperature, by computing
the volume of the static BTZ Janus black hole at time $t=0$.
In section \ref{sect:time-dependent-BTZ} we consider CV for the
time-dependent Janus BTZ black hole.
In this situation, the dual theory is not a defect CFT,  
but it corresponds to two entangled CFTs with different
values of the dilaton field on each side of the Penrose diagram.
We determine the time dependence of volume complexity for this situation.
We  summarize our results in section \ref{sect:conclu}. 
Appendix \ref{elliptics} contains our conventions
for Jacobi elliptic functions and elliptic integrals.

\section{Janus geometry}
\label{sect:janus_review}

In this section we introduce some of the faces that the Janus geometry can assume. The Janus solution is 
 a dilatonic domain-wall deformation of spacetime whose dual description is an interface CFT (ICFT).
We focus on the three-dimensional space, starting with the Janus deformation of empty $\mathrm{AdS}_3$ space described in section \ref{sect-Janus_AdS3}.
Complexity requires the introduction of UV regulators; these are naturally determined from the Fefferman-Graham (FG) expansion of the metric, which is considered in Section \ref{sect-FG_expansion}.
In Section \ref{sect-conformal_diagram_Janus} we present the conformal structure of the Janus AdS background, while in Section \ref{sect-static_JanusBTZ} we extend the static Janus deformation to the case of the BTZ black hole solution.

\subsection{Janus AdS$_3$}
\label{sect-Janus_AdS3}

The Janus AdS$_3$ solution \cite{Bak:2007jm} can be obtained from  type IIB
supergravity on $\mathrm{AdS}_3 \times S^3 \times M_4,$ where $M_4$ can be chosen
either as $T^4$ or $K3$. Upon dimensional reduction, the following action is obtained
\beq
I = \frac{1}{16 \pi G} \int d^3 x \sqrt{-g} \, \le R - \p^a \phi \p_a \phi + \frac{2}{L^2} \ri  \, ,
\eeq
where $\phi$ is the dilaton and $L$  the AdS$_3$ radius.

We will use the metric ansatz 
\beq
ds^2 = L^2 (f(y) ds^2_{\mathrm{AdS_2}} + dy^2) \, ,
\label{eq:Janus_metric}
\eeq
where $ds^2_{\mathrm{AdS_2}}$ denotes the generic AdS$_2$ metric in arbitrary coordinates.
Unless otherwise specified, we will describe the AdS slices using Poincarè coordinates, that is,
\beq
ds^2= L^2 \le f(y) \frac{d z^2-d t^2}{z^2} + dy^2  \ri \, .
\label{eq:Janus_metric_poincare}
\eeq
Solving the equation of motion, the following solution is obtained
\beq
\begin{aligned}
f(y) &= \frac{1}{2} \le 1 + \sqrt{1-2\gamma^2} \cosh (2y) \ri \, , \\
\phi(y) &= \phi_0 + \frac{1}{\sqrt{2}} \log \le \frac{1+\sqrt{1-2\gamma^2} + \sqrt{2} \gamma \tanh y}{1 + \sqrt{1-2\gamma^2}-\sqrt{2} \gamma \tanh y} \ri \, ,
\label{eq:function_f_and_dilaton}
\end{aligned}
\eeq
where the coordinate $y$ is ranged in $(-\infty,+\infty)$.
In correspondence of the extrema $y = \pm \infty$ we have two boundaries,
where the two sides of the interface field theory live. 
The parameter $\phi_0$ is set by the value of the dilaton at $y=0$. 
The constant $\gamma \in [0, \sqrt{2}/2]$
parameterizes the excursion of the dilaton between the two sides.
For $\gamma=0$ the dilaton is constant and the solution is just empty AdS.

The dual interpretation of this metric is a two-dimensional CFT deformed by a
 marginal operator $O(x)$  with couplings
 $J_\pm \int d^2x \, O(x)$ on each boundary,
 where
  \beq
  J_{\pm} = \lim_{y \rightarrow \pm \infty} \phi(y) \, .
  \eeq
 The bulk coordinate $y$
 connects two theories with different coupling constants
 living at $y=\pm \infty$.
For $\gamma=\sqrt{2}/2$, the dilaton is linear, i.e.,
$\phi=\phi_0+ \sqrt{2} y$ and $f$ is constant.
In this limit the dilaton is divergent at the boundary,
and the bulk theory is infinitely strongly coupled.
Since the Janus deformation is associated with an exactly marginal operator,
 it does not change the central charge of the CFT.

 For later purposes, it is useful to perform the change
 of variable $\mu(y)$ such that
 \beq
 d \mu= \frac{dy}{ \sqrt{f(y)}} \, .
 \label{change-vars-y-mu}
\eeq
 In the new set of coordinates the metric is
 \beq
ds_3^2 = L^2 f(y(\mu))  \cos^2 \mu \, ds^2_{\mathrm{AdS}_3} \, ,
\label{eq:metric_Janus_mucoord}
\qquad
ds^2_{\mathrm{AdS}_3} = \frac{1}{\cos^2 \mu} \le d\mu^2 + ds^2_{\mathrm{AdS}_2} \ri \, .
\eeq
The function $f(\mu)=f(y(\mu))$ and the dilaton can be expressed  \cite{Freedman:2003ax} in terms of Jacobi elliptic functions
(see Appendix \ref{elliptics} for our conventions)
\beq
\begin{aligned}
f (\mu) &= \frac{\alpha_+^2}{\mathrm{sn}^2 \le \alpha_+ (\mu + \mu_0), m \ri} \, ,  \\
\phi(\mu) &=\phi_0 + \sqrt{2} \log \left[  \mathrm{dn} \le \alpha_+ (\mu + \mu_0) , m \ri 
- \sqrt{m} \, \mathrm{cn} \le \alpha_+ (\mu + \mu_0),  m \ri \right] \, ,
\label{eq:solutions_f_dilaton_mu_coordinates}
\end{aligned}
\eeq
where 
\beq
\alpha_{\pm}^{2}=\frac{1}{2}\left(1 \pm \sqrt{1-2 \gamma^{2}}\right) \, , \qquad
m=\le \frac{\a_-}{\a_+} \ri^2 \, ,
\qquad
 \mu_0 =\frac{ \mathbb{K} ( m ) }{\alpha_+ }\, .
\eeq
The range of the new coordinate $\mu$ is $ [-\mu_0, \mu_0].$
Note that when $\gamma=0$ we get $\mu_0=\pi/2$
and we recover the AdS case with constant dilaton $\phi_0$ and
$f(\mu)=1/\cos^2 \mu$.

\subsection{UV regulator and Fefferman-Graham expansion}
\label{sect-FG_expansion}

\begin{figure}[h]
\begin{center}
\begin{tikzpicture}[thick,scale=1.1]

\draw [black,very thick,-stealth] (-5,0) -- (5,0);
\node (xaxis) at (4.5,0.2) {$\eta$};
\draw [black,very thick,-stealth] (0,0) -- (0,5);
\node (xaxis) at (0.2,4.5) {$\xi$};

\draw [qqwwzz,thick] (1,0) arc (0:180:1);
\draw [qqwwzz,thick](2,0) arc (0:180:2);
\draw [qqwwzz,thick](3,0) arc (0:180:3);
\node [qqwwzz] (label2) at (-1.7,3) {\textit{Constant} $z$};
\draw [qqwwzz,thick](4,0) arc (0:180:4);

\draw [ccqqqq,thick] (0,0) -- (-3,3);
\node [ccqqqq,thick] (label1) at (3.4,3.15) {\textit{Constant} $y$};
\draw [ccqqqq,thick](0,0) -- (3,3);

\draw [ccqqqq,thick](0,0) -- (-4,1);
\draw [ccqqqq,thick](0,0) -- (4,1);

\draw [ccqqqq,thick](0,0) -- (-1,4);
\draw [ccqqqq,thick](0,0) -- (1,4);
\draw [black,fill] (0,0) circle [radius=0.07]; 
\end{tikzpicture}
\caption{Foliation of the $\mathrm{AdS}_3$ geometry into $\mathrm{AdS}_2$ slices, 
where the lines at constant $y$ and $z$ coordinates are depicted. 
The black dot in the origin is the location of the Janus interface.}
\label{fig-AdS3_AdS2_slices}
\end{center}
\end{figure}
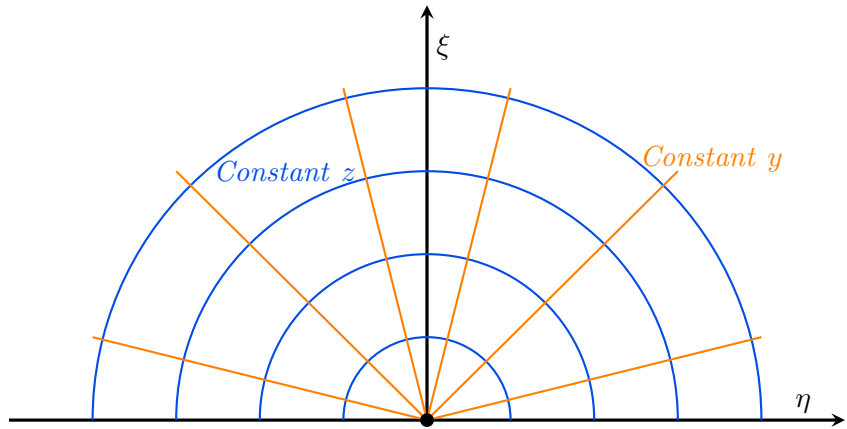

A standard procedure for regularizing AdS-sliced metrics is to introduce a constant cutoff 
along the radial direction in a Fefferman-Graham (FG) expansion of the metric
 (see, e.g.,  \cite{Papadimitriou:2004rz} and \cite{Chiodaroli:2010ur}).
  Let's start by considering the $\text{AdS}_2$-sliced metric
  for the pure $\text{AdS}_3$ geometry, 
 \beq
ds^2_{\mathrm{pure}} 
= L^2 \left( d y^2 + \cosh^2 y \, \frac{d z^2-d t^2}{z^2} \right)  \, ,
\label{eq:AdS_2_slicing}
\eeq 
that is, the Janus geometry with $\gamma=0$. The boundary parameterization of $\mathrm{AdS}_3$ in the coordinates (\ref{eq:AdS_2_slicing}) is more complicated compared to the one in Poincarè coordinates. In fact, the two $(1+1)-$dimensional boundaries at $ y \rightarrow\pm\infty$  in \eqref{eq:AdS_2_slicing} sit on the same boundary when represented in Poincarè coordinates. Let us first relate the metric \eqref{eq:AdS_2_slicing} to the same metric in Poincaré coordinates
\beq
\label{eq:AdS_3_pure}
ds^2_{\text{pure}} = \frac{L^2}{\xi^2}\left(d\xi^2+d\eta^2-dt^2\right) \, ,
\eeq
by implementing the change of variables (see fig.~\ref{fig-AdS3_AdS2_slices})
\beq
\eta= z \tanh y \, , \qquad
\xi= \frac{z}{\cosh y} \, .
\label{eq:change_coordinates_1_FG}
\eeq

Coming back to the Janus geometry in Eq.~(\ref{eq:Janus_metric}),
 we can see that in the boundary regions $y\rightarrow\pm\infty$ the metric 
 can be approximated by 
\begin{equation}
	ds^2_{\pm\infty}=L^2\left(dy^2+\frac{\sqrt{1-2\gamma^2}}{4}e^{\pm 2 	y}\frac{dz^2-dt^2}{z^2}	\right).
	\label{eq-feffi}
\end{equation}
Equation (\ref{eq-feffi}) simplifies by shifting the $y$ coordinate as follows
\begin{align}
	\tilde{y}=y\pm\frac{1}{2}\log \sqrt{1-2\gamma^2}, \quad   ds^2_{\pm		\infty}=L^2\left(d\tilde{y}^2+\frac{1}{4}e^{\pm 2 \tilde{y}}			\frac{dz^2-dt^2}{z^2}	\right).
	\label{eq:change_variable_ytilde}
\end{align}
We can then introduce the FG coordinates $(\xi,\eta)$: 
\begin{itemize}
\item when $\tilde{y}\rightarrow\infty$, $\xi\rightarrow 0$, and $\eta> 0$ 
\begin{align}
	e^{-2\tilde{y}}=\frac{1}{4}\frac{\xi^2}{\eta^2}, \qquad z=\eta 			\left(1 + \frac{1}{2}\frac{\xi^2}{\eta^2}\right) \, ,
	\label{eq:change_coordinates_FG_2}	
\end{align}
\item  when $\tilde{y}\rightarrow-\infty$, $\xi\rightarrow 0$, and $\eta<0$ 
\begin{align}
	e^{2\tilde{y}}=\frac{1}{4}\frac{\xi^2}{\eta^2}, \qquad z=|\eta| 			\left(1 +  \frac{1}{2}\frac{\xi^2}{\eta^2}\right) .
	\label{eq:change_coordinates_FG_3}	
\end{align}
\end{itemize}
Therefore, the metric in Eq.~(\ref{eq-feffi}) becomes
\begin{align}
	ds^2_{\pm\infty}=\frac{L^2}{\xi^2}\left(d\xi^2+d\eta^2-dt^2 + \mathcal{O} (\xi) \right).
\end{align}
This is valid when we are far from the interface located at $(\eta,\xi)=0$.
In fact, close to the origin the coordinate transformation 
is singular and a different cutoff should be introduced. More precisely, the regime where the FG expansion is valid 
is $\xi  \ll \eta, $ which implies that at the lowest order $|\eta| \simeq z $, see eqs. \eqref{eq:change_coordinates_FG_2}
and \eqref{eq:change_coordinates_FG_3}. We will discuss how to regularize divergences in the region close to the interface in section \ref{sect-different_regularizations}.

\subsection{Conformal diagram of the Janus geometry}
\label{sect-conformal_diagram_Janus}

\begin{figure}[h]
\begin{center}
\begin{tikzpicture}[thick,scale=1.2]
\draw (-2,2) -- (0,0)
node[midway, left, inner sep=2mm] {$-\mu_0$};
\draw (2,2) -- (0,0)
node[midway, right, inner sep=2mm] {$\mu_0$};
\draw [dashed] (0,0) -- (0,-1.5);
\node (w) at (0,0.3) {$W$};

\end{tikzpicture}
\caption{Conformal diagram taken from \cite{Bak:2003jk} for the Janus $\mathrm{AdS}_3$ geometry with Poincaré coordinates on the $\mathrm{AdS}_2$ slices. The polar angle corresponds to the coordinate $\mu \in [-\mu_0, \mu_0]$ and the radial coordinate is $z\in [0, \infty].$ The joint $W$ corresponds to the place where the two parts with the topology of half $\mathbb{R}^2$ meet. It can be seen as a domain wall on the boundary.}
\label{fig-conformal_diagram_Janus_AdS}
\end{center}
\end{figure}
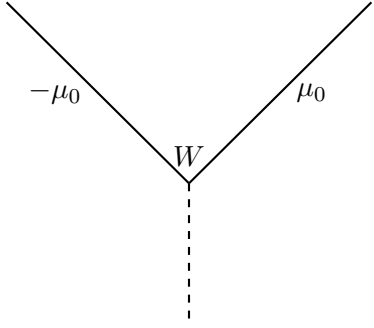

In order to understand the geometrical properties of Janus,
it is useful to consider the conformal diagram.
It can be obtained by identifying an appropriate conformal factor multiplying a known metric.
Starting from Eq.~(\ref{eq:metric_Janus_mucoord}), and choosing
the $\mathrm{AdS}_2$ slices in Poincaré coordinates, we obtain 
\beq
ds^2_3 = \frac{L^2}{z^2} \, f(\mu) \le z^2 d\mu^2 + d z^2 - dt^2 \ri \, .
\eeq 
This metric is conformal to a portion of the Minkowski spacetime
 where the spatial part is written in polar coordinates with angle $-\mu_0 \leq \mu \leq \mu_0$ and radius $z>0.$ 
This conformal diagram is depicted in Fig.~\ref{fig-conformal_diagram_Janus_AdS}.
Since $\mu_0 \geq \pi/2, $ the junction between the half-boundaries meets at a joint\footnote{While this joint seems singular, it can be shown that it is only an artifact of the coordinate system.} $W$ with an obtuse angle.

\begin{figure}[h]
\begin{center}
\begin{tikzpicture}[thick,scale=1.2]
\draw[dashed] (0,-2) -- (0,2);
  \draw (0,0) circle (2cm);
  \draw[very thick] (0,2) arc (53.130089864:-53.130089864:2.5);
  
    \draw[very thick] (0,-2) arc (53.130089864:-53.130089864:-2.5);
\node (a) at (-1.3,0) {$-\mu_0$}; 
\node (b) at (1.3,0) {$\mu_0$}; 
  \fill[fill=black] (0,2) circle (1pt) node [above] {$N$};
   \fill[fill=black] (0,-2) circle (1pt) node [below] {$S$};
\end{tikzpicture}
\caption{Conformal diagram for the Janus $\mathrm{AdS}_3$ geometry with global coordinates on the $\mathrm{AdS}_2$ slices.}
\label{fig-conformal_diagram_Janus_AdS_global}
\end{center}
\end{figure}
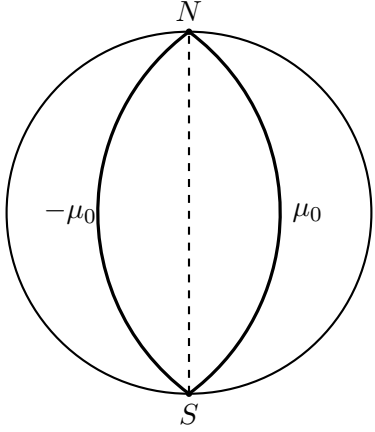

We can also consider the case where global coordinates are taken on AdS$_2$ the slicing,
which gives the metric
\beq
ds^2 = \frac{L^2}{\cos^2 \lambda} \le -dt^2 + \cos^2 \lambda \, d\mu^2 + d\lambda^2 \ri \, ,
\eeq
and the associated conformal diagram is given in Fig.~\ref{fig-conformal_diagram_Janus_AdS_global}.
The boundary consists of two parts, defined by $\mu = \pm \mu_0,$ which are 
given by two halves of $S^1$ joined through the north and south poles.

\subsection{Static Janus BTZ black hole}
\label{sect-static_JanusBTZ}

The static Janus BTZ background \cite{Bak:2011ga, Bak:2020enw} 
 is dual to the Janus interface conformal field theory at finite temperature.
The Kruskal extension of this solution is dual to two copies of the 
Janus interface CFT in the thermofield double state.
This solution has a timelike Killing vector, hence the name static. It can be obtained as follows.

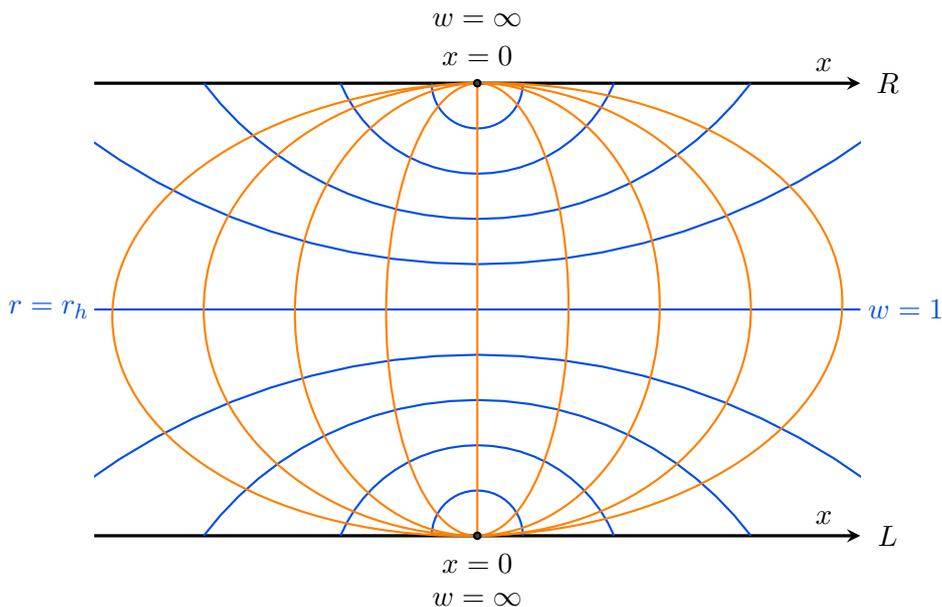
\begin{figure}[ht]
\begin{center}
\begin{tikzpicture}[thick,scale=1.2]
\node [color=qqwwzz] at (4.7,0) {$w=1$};
\node [color=qqwwzz] at (-4.7,0) {$r=r_h$};
\node  at (4.5,2.5) {$R$};
\node  at (0,2.8) {$x=0$};
\node  at (0,3.2) {$w=\infty$};
\node  at (0,-3.2) {$w=\infty$};
\node  at (0,-2.8) {$x=0$};
\node  at (4.5,-2.5) {$L$};
\node at (3.8,2.7) {$x$};
\node at (3.8,-2.3) {$x$};
\clip(-4.2,-3) rectangle (4.2,3);
\draw [very thick,-stealth] (-4.2,2.5)-- (4.2,2.5);

\draw  [very thick,stealth-] (4.2,-2.5)-- (-4.2,-2.5);

\draw [shift={(0,-2.5)},color=qqwwzz]  plot[domain=0:3.141592653589793,variable=\t]({1*0.5*cos(\t r)+0*0.5*sin(\t r)},{0*0.5*cos(\t r)+1*0.5*sin(\t r)});
\draw [shift={(0,-3.125)},color=qqwwzz]  plot[domain=0.39479111969976155:2.746801533890032,variable=\t]({1*1.625*cos(\t r)+0*1.625*sin(\t r)},{0*1.625*cos(\t r)+1*1.625*sin(\t r)});
\draw [shift={(0,-4.75)},color=qqwwzz]  plot[domain=0.6435011087932844:2.498091544796509,variable=\t]({1*3.75*cos(\t r)+0*3.75*sin(\t r)},{0*3.75*cos(\t r)+1*3.75*sin(\t r)});
\draw [shift={(0,-7.75)},color=qqwwzz]  plot[domain=0.8097835725701669:2.3318090810196264,variable=\t]({1*7.25*cos(\t r)+0*7.25*sin(\t r)},{0*7.25*cos(\t r)+1*7.25*sin(\t r)});
\draw [shift={(0,2.5)},color=qqwwzz]  plot[domain=-3.141592653589793:0,variable=\t]({1*0.5*cos(\t r)+0*0.5*sin(\t r)},{0*0.5*cos(\t r)+1*0.5*sin(\t r)});
\draw [shift={(0,3.125)},color=qqwwzz]  plot[domain=3.5363837732895544:5.8883941874798245,variable=\t]({1*1.625*cos(\t r)+0*1.625*sin(\t r)},{0*1.625*cos(\t r)+1*1.625*sin(\t r)});
\draw [shift={(0,4.75)},color=qqwwzz]  plot[domain=3.7850937623830774:5.639684198386302,variable=\t]({1*3.75*cos(\t r)+0*3.75*sin(\t r)},{0*3.75*cos(\t r)+1*3.75*sin(\t r)});
\draw [shift={(0,7.75)},color=qqwwzz]  plot[domain=3.95137622615996:5.4734017346094195,variable=\t]({1*7.25*cos(\t r)+0*7.25*sin(\t r)},{0*7.25*cos(\t r)+1*7.25*sin(\t r)});
\draw [color=qqwwzz] (-4.2,0)-- (4.2,0);

\draw [rotate around={-90.36548428604479:(0,0)},color=ccqqqq] (0,0) ellipse (2.5002670917768683cm and 0.9999829093034179cm);
\draw [rotate around={-91.94780675769272:(0,0)},color=ccqqqq] (0,0) ellipse (2.500814570209908cm and 1.999583274080338cm);
\draw [rotate around={-179.7215782473841:(0,0)},color=ccqqqq] (0,0) ellipse (3.0000155856698854cm and 2.4999909806342093cm);
\draw [rotate around={-178.10462974282584:(0,0)},color=ccqqqq] (0,0) ellipse (4.00342489873214cm and 2.499165333952163cm);
\draw [color=ccqqqq] (0,2.5)-- (0,-2.5);
\begin{scriptsize}
\draw [fill=uuuuuu] (0,2.5) circle (1pt);
\draw [fill=uuuuuu] (0,-2.5) circle (1pt);
\end{scriptsize}
\end{tikzpicture}
\end{center}
\caption{Picture representing an AdS slice of the BTZ black hole at constant time. 
The lines at constant $w$ are depicted in blue, while the curves at constant $\mu$ in orange.
 The coordinate $x$ runs along the left ($L$) and right ($R$) disconnected boundaries. 
The coordinate $r$ covers the region outside the horizon and runs on the vertical axis: starting from the middle line located at $r=r_h,$ it increases towards the L and R boundaries. }
\label{fig-BTZ_AdS_slicing}
\end{figure}

Starting from the Janus AdS metric
\beq
ds^2 = L^2 \le dy^2 + f(y) \, ds^2_{\mathrm{AdS}_2} \ri
= L^2 \, f(\mu) \le d\mu^2 + ds^2_{\mathrm{AdS}_2} \ri \, ,
\label{eq:general_form_3d_Janus}
\eeq
one can replace Poincaré  AdS$_2$ with Rindler AdS
\beq
ds^2_{\mathrm{AdS}_2} \rightarrow ds^2_{\rm Rindler} = - \frac{(w^2-1) \, r_h^2}{L^4} dt^2 + \frac{dw^2}{w^2-1} \, .
\label{eq:Rindler_slicing}
\eeq
For $\gamma=0$,  the interfaces on each boundary disappear and
the solution coincides with the  BTZ black hole \cite{Banados:1992gq}.
This can be seen from the change of coordinates (see figure \ref{fig-BTZ_AdS_slicing})
\beq
\frac{r_h}{r} = \frac{\cos \mu}{\sqrt{w^2 - \sin^2 \mu}} \, , \qquad
\sinh \frac{r_h x}{L^2} = \frac{\sin \mu}{\sqrt{w^2 - \sin^2 \mu}} \, ,
\eeq
which bring the metric in Eqs.~(\ref{eq:general_form_3d_Janus}) and (\ref{eq:Rindler_slicing}) to the
familiar form of the BTZ black hole  with vanishing angular momentum
\beq
ds^2_{\mathrm{BTZ}} = - \frac{r^2 - r_h^2}{L^2} dt^2 + \frac{L^2 dr^2}{r^2 - r_h^2} + \frac{r^2}{L^2} dx^2 \, .
\label{eq:BTZ_metric}
\eeq
 The horizon is at $r = r_h$ and $r = 0$ corresponds to an orbifold singularity.
 This solution is invariant under time $t$ and spatial $x$ translations. When the deformation parameter $\gamma$ is non vanishing, translational 
 invariance in $x$ is broken, while the time translation symmetry in $t$ remains unbroken.
 When $\gamma=0$, due to translational invariance in $x$, we can compactify
 $x \sim x+2 \pi$. For general $\gamma$, this identification is no longer 
 consistent with the symmetries of the problem.
 
For generic $\gamma$
the asymptotic behaviour of the solution can be mapped to the asymptotics of the 
BTZ as follows
\beq
\label{eq:asymp_JBTZ}
\frac{r}{r_h} \simeq \sqrt{(w^2-1)f(y)+1} \, , \qquad
\sinh \frac{r_h x}{L^2} \simeq \mathrm{sign}(x) \, \sqrt{\frac{f(y)-1}{(w^2-1)f(y)+1}} \, .
\eeq
At generic $\gamma$, the solution has a Killing vector $\partial_t$,
which corresponds to time translations.
When considering the Kruskal extension of the solution,
this Killing vector corresponds to the following time translations
on the left and right boundaries
\beq
t_L \rightarrow t_L \pm \Delta t \, , \qquad
t_R \rightarrow t_R \mp \Delta t \, ,
\label{bountary-time-shift}
\eeq
The field theory dual involves the left and the right interface CFTs 
in the thermofield (TFD) double state
\beq
| \psi (t_L, t_R) \rangle = e^{-i \le t_L \, H \otimes \mathbf{1} + t_R \, \mathbf{1} \otimes H \ri} | \psi(0,0) \rangle \, , 
\qquad
| \psi(0,0) \rangle  = \frac{1}{\sqrt{Z}} \sum_n e^{-\frac{\beta}{2} E_n} |n \rangle \otimes | n \rangle \, .
\label{eq:TFD_Janus_BTZ_static}
\eeq
The choice $t_L = -t_R$ is time-independent,
and corresponds to the $\partial_t$ Killing vector of the geometry.

\section{Volume for Janus AdS$_3$}
\label{sect-CV_Janus}

In this section we apply the CV conjecture to the Janus AdS$_3$ geometry.
We consider an extremal spacelike codimension-one slice attached to the boundary and we evaluate the induced volume.
Since this geometric object extends all the way to the boundary, 
the corresponding holographic complexity will be divergent and it is necessary to regularize the UV modes.
In the case of entanglement entropy, divergences
arise due to the arbitrarily short correlations between degrees of freedom 
on each side of the entangling surface.
The leading divergence scales with the area law and either the finite term
 (in odd spacetime dimensions) or the coefficient of the logarithmic divergence
  (in even spacetime dimensions) have a universal interpretation which is 
  not sensitive to the ambiguities in the choice of the regulator.
In the case of complexity, the outcome of a similar classification is a leading divergence proportional to the boundary volume of the time slice, 
 and  a set of subleading terms 
 defined in terms of  integrals over the same slice \cite{Carmi:2016wjl}.

However, this structure may change when defects, interfaces, or boundaries are present, 
and it is interesting to understand how the structure gets modified.   
This problem has already been addressed in some specific cases
\cite{Chapman:2018bqj,Sato:2019kik, Braccia:2019xxi}.
In this section we will study the UV divergences of the AdS$_3$
Janus geometry. This analysis will clarify the universality of UV divegences in spacetimes
 containing a codimension-one object
 by comparing with the Randall-Sundrum model \cite{Chapman:2018bqj} or
  with the case of a boundary CFT holographically described by an end of the world brane \cite{Sato:2019kik, Braccia:2019xxi}.

In the calculation of the volume, an IR regulator is also needed in order to get a finite result.
To deal with this divergence, a possibility could be to introduce an IR cutoff at some constant value of the Fefferman-Graham (FG) coordinates.
This approach is not very practical for the Janus geometry, as the FG coordinates
are known just as an expansion nearby the boundary. 
In our case, it is more convenient to use
the related concept of subregion complexity  as an IR regulator.

Subregion volume complexity \cite{Alishahiha:2015rta} is from the bulk perspective defined as the volume of an extremal slice
limited by a constant time subregion on the boundary and the corresponding  HRT surface.
The precise meaning of subregion complexity in the dual CFT is still an open question.
Some proposals, such as fidelity, purification complexity and basis complexity,
were discussed in \cite{Alishahiha:2015rta,Agon:2018zso}. 
 In any event, the HRT surface defining the subregion volume can effectively be used as an infrared regulator:
  the total complexity will then be defined as the limit in which the subregion
   covers the entire boundary. In the case of the Janus geometries under consideration in this section,
    the spacetime is $(2+1)$-dimensional and then the RT surface is a geodesic.

In section \ref{sect-time_dependence_Janus_AdS}, we discuss the time independence of the volume 
of the Janus AdS geometry.
In section \ref{sect-different_regularizations} 
 three different prescriptions to regulate the UV divergences of the extremal volume
 are described;  these regularization only differ by finite terms.
 In section \ref{sect-Janus_one_cutoff} we  compute  
 the Janus AdS$_3$ volume, using the single cutoff regularization;
 in sections \ref{app-static_volume_FG_method} and 
 \ref{app-Janus_AdS3_double_cutoff} we will discuss how the computation is modified
 using  the FG and the double cutoff regularizations.
We will show that these regularization prescriptions differ only by a finite term.
In section \ref{comparison-other-geometries} we will compare the result
with the one found in other AdS defect geometries.

\subsection{Time independence of the volume in the Janus $\mathrm{AdS}_3$ geometry}
\label{sect-time_dependence_Janus_AdS}

We want to show that it is not restrictive to study the CV conjecture in 
the Janus $\mathrm{AdS}_3$ background using a time slice at constant boundary time.
One could be tempted to assign different time arrows on each of the two $y=\pm \infty$
boundaries (as customary for the left and right side of the Kruskal diagram).
 However, this is not consistent because the two boundaries are not causally disconnected.
 We are then forced to evolve the time in a unique way
  according to the asymptotic Killing vector of the metric.
  This forces us to take the boundary condition of the extremal volume at constant time $t$. 
  The discussion also applies to the HRT surface
  associated to the subregion, which is also time independent.

  Let us now show that the whole solution is at constant $t$.
   We parametrize the codimension-one slice expressing the time as a function  $t(y,z).$
From the metric in Eq.~(\ref{eq:Janus_metric_poincare}), the volume functional is
\beq
\mathcal{V} = L^2 \int dz \, dy \, \mathcal{L} \, ,
\qquad
 \mathcal{L} = \frac{\sqrt{f(y)}}{z} \, \sqrt{ 1- \le \p_y t  \ri^2 - \frac{f(y)}{z^2} \,  \le \p_z t  \ri^2 } \, .
 \label{eq:Lagrangian_Janus_AdS3_time}
\eeq
The $t=t_0$ function, where $t_0$ is a constant, is  a solution of the Euler-Lagrange equation.
Moreover, this solution has the property to maximize the volume functional.

\subsection{UV regularizations for  the volume}
\label{sect-different_regularizations}

The volume is divergent and needs a regularization by an UV cutoff.
The choice of this cutoff is a delicate issue in Janus geometries,
due to the presence of subtleties with the Fefferman-Graham (FG) coordinates.
Nearby the defect, there is a region where the FG patch is not well-defined
and the UV cutoff  surface is ambiguous. 
To overcome this problem, we will use  three different regularization prescriptions
and we will show that they only differ by finite terms.
Each regularization is equally valid to describe the relevant physics of the system.

We generically consider an interface CFT described
 by a codimension-one defect embedded in a $\mathrm{AdS}_{d+1}$ bulk geometry, 
 whose isometries get reduced from the conformal group $\mathrm{SO}(d,2)$ to the subgroup $\mathrm{SO}(d-1,2).$
The natural way to parametrize this geometry is to perform a slicing of spacetime in terms of $\mathrm{AdS}_d$ slices.
The metric takes the form \cite{Estes:2014hka}
\beq
ds^2 = L^2 \le  A^2 (y) ds^2_{\mathrm{AdS}_d} + \rho^2(y) dy^2 \ri \, ,
\label{eq:metric_Trivella_form}
\eeq
with $y$ being a non-compact coordinate such that when $y \rightarrow \pm \infty$ we have the asymptotic behaviour
\beq
A(y) \rightarrow \frac{L_{\pm}}{2} e^{\pm y \pm c_{\pm}} \, , \qquad
\rho(y) \rightarrow 1 \, .
\label{eq:asymptotic_Trivella_functions}
\eeq
Here $L_{\pm}$ and $c_{\pm}$ are constants (which can take different values at the boundaries $y= \pm \infty$), and we are assuming that there is not any other internal direction in the spacetime.
We parametrize the $\mathrm{AdS}_d$ slices using Poincaré coordinates 
\beq
ds^2_{\mathrm{AdS}_d} = \frac{1}{z^2} \le dz^2 - dt^2 + d \vec{x}^2_{d-2} \ri \, ,
\label{eq:metric_AdS_slicing}
\eeq
where $(t,z)$ are the time and radial coordinates on each slice and $\vec{x}$ collects all the other orthogonal directions. 
In the following, we will use three regularization techniques inspired 
by the similar discussion for the free energy \cite{Bak:2016rpn} and the entanglement entropy  \cite{Estes:2014hka,Gutperle:2016gfe}.
In all the computations of this Section, the volume will be determined from
\beq
\mathcal{V} = \int dz \int dy \int d\vec{x} \, \sqrt{h} \, ,
\label{eq:general_prescription_volume}
\eeq
where $\sqrt{h}$ is the determinant of the induced metric.
The integration along the orthogonal spatial directions $\vec{x}$ is usually trivial, 
while the part along the $(y,z)$ coordinates contains the relevant information about the defect. 

The three different regularization that we will consider are:
\begin{itemize}
\item
\textbf{Fefferman-Graham regularization.} 
The Fefferman-Graham (FG) form of the metric is
\beq
ds^2 = \frac{L^2}{\xi^2} \left[ d \xi^2 + g_1 (\xi/\eta) \,  \le -dt^2 + d\vec{x}^2 \ri + g_2 (\xi/\eta) \, d\eta^2  \right] \, ,
\label{eq:form_metric_FGexpansion}
\eeq
where $\xi$ is a radial coordinate for the asymptotic AdS region in Poincaré coordinates,
 $\eta$ is a field theory direction orthogonal to the defect, and $g_1, g_2$ are two opportune functions,
 such that  the original metric \eqref{eq:metric_Trivella_form} with slicing \eqref{eq:metric_AdS_slicing}
 is equivalent to (\ref{eq:form_metric_FGexpansion}) with a suitable change of coordinates $(z,y) \rightarrow (\xi, \eta).$ 
We perform an expansion of Eq.~\eqref{eq:form_metric_FGexpansion} such that the asymptotic metric reads
\beq
ds^2 = \frac{L^2}{\xi^2} \le d\xi^2 + d\eta^2 + d\vec{x}^2 - dt^2 + \mathcal{O}(\xi) \ri \, .
\label{eq:metric_FG_Poincare_expansion}
\eeq
The natural prescription to regularize divergences using the FG form of the metric is to
 introduce a UV cutoff by cutting the spacetime with the surface located at $\xi= \delta,$ and expand all the results in a series around $\delta =0.$
The problem of this procedure is that in the region where $\xi/\eta \gg 1,$ 
 the FG expansion breaks down and the coordinates $(\xi, \eta)$ are ill-defined \cite{Papadimitriou:2004rz}.

For this reason, the defect geometry is characterized by the existence of two patches, defined away from the region of the defect on the left and right sides of the spacetime, where the FG expansion is valid: we call them FG patches (see figure \ref{fig-interpolation_FG_patches}.)
We do not have access to a natural 
 UV cutoff in the middle region closer to the defect.
 To overcome the problem,
the original proposal from \cite{Estes:2014hka} is
 to interpolate the cutoff determined by requiring $\xi=\delta$
  in the left and right FG patches with an arbitrary curve in the middle region.
   The only constraint is that the curve should be continuous at the value $y=y_0$ where the FG expansion breaks down.
The corresponding curve is pictorially represented in Fig.~\ref{fig-interpolation_FG_patches}.

\begin{figure}[h]
\begin{center}
\begin{tikzpicture}[thick,scale=1.3]
\fill[red!40!yellow] (0,0) -- (-0.75,3) -- (0.75,3) -- (0,0);
\fill[black!20!white] (0,0) -- (-5,0) -- (-5,3) -- (-0.75,3) -- (0,0);
\fill[black!20!white] (0,0) -- (5,0) -- (5,3) -- (0.75,3) -- (0,0);
\draw (-5,0) -- (5,0)
node[midway, below, inner sep=2mm] {\textit{Defect}};
\draw (-5,1) -- (-0.26,1)
node[midway, below, inner sep=1mm] {$\xi=\delta$}
node[midway, above, inner sep=9mm] {\textit{Left FG Patch}};
\draw (0.26,1) -- (5,1) 
node[midway, below, inner sep=1mm] {$\xi=\delta$}
node[midway, above, inner sep=9mm] {\textit{Right FG Patch}};
\draw (0.26,1) arc (45:135:0.37)
node[midway, above, inner sep=1mm] {\textit{$\Gamma$}};

\draw (0,0) -- (-0.75,3);
\draw (0,0) -- (0.75,3);
\draw [black,fill] (0,0) circle [radius=0.07]; 
\end{tikzpicture}
\caption{Interpolation between two FG patches with a continuous curve $\Gamma$.
}
\label{fig-interpolation_FG_patches}
\end{center}
\end{figure}
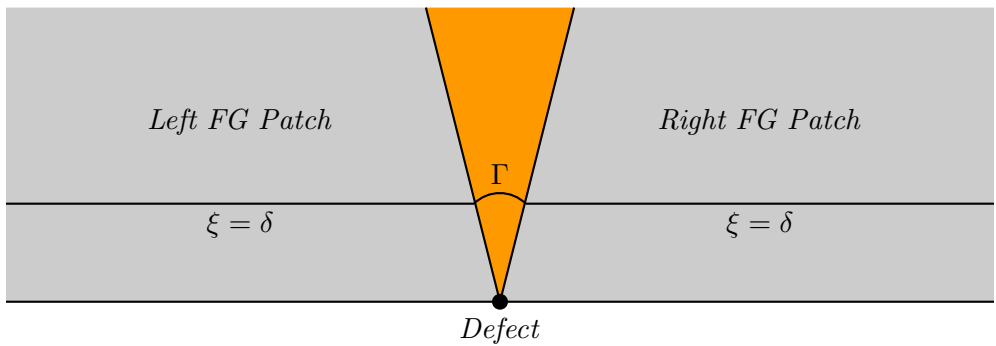

This method was later applied to the calculation of complexity
 in \cite{Chapman:2018bqj}, with the additional requirement that the interpolation is smooth,
  i.e., the curves in the middle region are perpendicular to the surface which delimits the FG patches.
We will discuss the details of this method in
section \ref{app-static_volume_FG_method}.

\item \textbf{Single cutoff regularization.} 
This technique \cite{Bak:2016rpn}  is inspired by the Fefferman-Graham method, 
but has the advantage to not introduce any arbitrary interpolating curve in the middle region.
 Instead, it uses the FG map to induce a minimal value on the $z$ coordinate,
  in such a way that the integration does not reach the region $z \rightarrow 0$ where the expansion breaks down.
We explain how the procedure works starting from  $\mathrm{AdS}_{d+1}$, which can be written using the slicing \eqref{eq:metric_Trivella_form} 
choosing $A(y)=\cosh y$ and $\rho(y)=1.$
In this case, the coordinate transformation which brings the metric to the FG 
form in eq.~\eqref{eq:form_metric_FGexpansion} is
\beq
\eta= z \tanh y \, , \qquad \xi = \frac{z}{\cosh y} \, .
\eeq
If we locate the UV cutoff at the surface $\xi=\delta,$ we get the condition
\beq
\delta = \frac{z}{\cosh y}  \, ,
\label{eq:one_cutoff_FG_empty_AdS}
\eeq
which selects a maximal value of $y=y^*(z)$
 for the first non-trivial integration in Eq.~\eqref{eq:general_prescription_volume}.
Reversing this formula gives a constraint on the minimal value of the integration along $z,$ determined by
\beq
z_{\rm min} = \delta \,  \underset{y \in \mathbb{R}}{\mathrm{min}} \le  \cosh y \ri = \delta  \, .
\eeq
In this way, we observe that the choice of a single cutoff $\delta$ from the FG expansion restricts the integration along both the $(y,z)$ coordinates and regularizes the volume.

In the presence of a defect, the procedure is the same, except that 
the conditions determined from the FG form of the metric get modified to
\beq
 \delta=\frac{z}{A(y)}  \, , \qquad
z_{\rm min} = \delta \, \underset{y \in \mathbb{R}}{\mathrm{min}} \, [A(y)] \, .
\label{eq:single_cutoff_prescription}
\eeq
We apply this technique in section \ref{sect-Janus_one_cutoff}.


\item \textbf{Double cutoff regularization.}
The previous technique regularizes all the integrals with the choice of a single UV cutoff inspired by the FG expansion.
On the other hand, we can consider  two different cutoffs for
the each of the directions  $(y,z)$.
This method  \cite{Gutperle:2016gfe} is based on the observation that, after the subtraction of the vacuum geometry, 
we should obtain a holographic quantity intrinsic to the defect: for this reason, 
a natural cutoff can be imposed on the $\mathrm{AdS}_d$ slicing at $z= \delta,$ instead
 of selecting the asymptotic radial direction in the $\mathrm{AdS}_{d+1}$ bulk geometry.
This choice by itself is not sufficient to regularize the full integral \eqref{eq:general_prescription_volume}, since the metric factor $A(y)$ is still singular at infinity: for this reason we also determine a maximum value of $y$ where the integration ends by requiring
\beq
A(y) = \frac{1}{\varepsilon} \, .
\label{eq:double_cutoff_prescription} 
\eeq 
Notice that while the $\delta$ cutoff has physical relevance as it regularizes the intrinsic contribution from the defect, the $\varepsilon$ cutoff is a mathematical artifact introduced at intermediate steps, and the result should therefore be $\varepsilon$-independent. As a consequence of this, we are allowed to remove the $\varepsilon$ cutoff at the end of the computation.
We apply this procedure in section \ref{app-Janus_AdS3_double_cutoff}.
\end{itemize}

\subsection{Single cutoff regularization}
\label{sect-Janus_one_cutoff}

We will consider a symmetric region of radius $l/2$ centered on the defect
in  Janus AdS$_3$ and we will compute its subregion complexity.
The total complexity can be defined as the $l \rightarrow \infty$ limit
of this result, with the length $l$ playing the role of IR regulator.

The interval is located at the FG radial coordinate $\xi=0$ and placed symmetrically
 along the orthogonal direction to the interface, i.e., $\eta \in [-\frac{l}{2}, \frac{l}{2}].$ 
There is an ambiguity in the regularization of the UV divergences: we can either put the subregion 
on the cutoff surface $\xi = \delta$ and build the corresponding HRT surface, or we can put the interval 
on the real boundary $\xi=0$ and then cut the HRT surface with the line at $\xi=\delta.$ 
Since the difference between the two cases vanishes in the $\delta \rightarrow 0  $ limit, we will only focus on the latter case.

As pointed out in \cite{Gutperle:2016gfe}, on a fixed time slice there is a particular 
simple class of geodesics for the Janus geometry given by curves at constant $z.$ 
Using the FG expansion in Eq.~\eqref{eq:change_coordinates_FG_2} and 
putting the boundary conditions $\xi=0$ and $\eta=\frac{l}{2},$ we determine the constant $\bar{z}$ value where the surface is located:
\beq
\bar{z} = \frac{l}{2}  \, .
\label{eq:definition_zbar}
\eeq
The prescription for the extremal volume tells us to consider a solution at $t=0$ anchored at the boundary and delimited by the HRT surface.

The UV divergencies will be regularised according to the single cutoff prescription
described in Section \ref{sect-different_regularizations}.
The correspondence between the generic form \eqref{eq:metric_Trivella_form}
 of the metric with a conformal defect and the Janus background is
\beq
A^2 (y) = f(y) \, , \qquad
\rho(y) = 1 \, , \qquad
L_{\pm}^2 = \sqrt{1-2\gamma^2} \, , \qquad
c_{\pm} = 0 \, .
\label{eq:dictionary_Janus_Trivella}
\eeq
In this way the requirement in Eq.~\eqref{eq:single_cutoff_prescription}
 to regularize the UV divergences becomes
\beq
\frac{z}{\sqrt{f(y)}} = \delta  \, .
\eeq
This single condition identifies both a maximum value of the coordinate $y=y^*(z),$ obtained by inverting the previous expression, and a minimum value of the coordinate $z=z_{\rm min},$ determined from the second identity in Eq.~\eqref{eq:single_cutoff_prescription}.
Concretely, they are given by
\beq
y^*(z) = f^{-1} \le \frac{z^2}{\d^2} \ri 
  \, , \qquad
z_{\rm min} = a \, \delta \, , \qquad a^2=f(0) =  \frac{1+\sqrt{1-2\gamma^2}}{2 }\, .
\label{eq:ystar_Janus_one_cutoff}
\eeq
The integral which computes the subregion volume in
 Eq.~\eqref{eq:general_prescription_volume} takes the form
\beq
\mathcal{V} (l,\gamma) = 2 L^2 \int_{z_{\rm min}}^{\bar{z} } \frac{dz}{z} \int_0^{y^*(z)} dy \, \sqrt{f(y)} \, .
\label{eq:integral_one_cutoff}
\eeq
Changing variables into 
\beq
 \tau=\frac{f(y)}{a^2} \, , \qquad \zeta=\frac{z^2}{a^2 \d^2} \, , \qquad
 \bar{\zeta}= \frac{\bar{z}^2}{a^2 \d^2} \, ,
\eeq
 we can express the integral as
\beq
\begin{aligned}
\mathcal{V} (l,\gamma) &=\frac{ L^2  a^3}{ \sqrt{2}} \int_{1}^{\bar{\zeta} } \frac{d \zeta}{\zeta}
 \int_{1}^{\zeta} \frac{\tau^{1/2} d \tau}{\sqrt{\gamma^2 +2 a^2 \tau (a^2 \tau-1) }}  \\
 &= \frac{ L^2  a^3}{ \sqrt{2}}
 \int_1^{\bar{\zeta}} d \tau \int_\tau^{\bar{\zeta}} \frac{d \zeta}{\zeta} 
 \frac{\tau^{1/2} }{\sqrt{\gamma^2 +2 a^2 \tau (a^2 \tau-1) }} \, .
\end{aligned}
\eeq
It is useful to define
\beq
m = \frac{1 - \sqrt{1-2\gamma^2}}{1 + \sqrt{1-2\gamma^2}} \, ,
\label{eq:a_and_k_definitions}
\eeq
which is one of the zeroes of the denominator.
Since $0 \leq \gamma \leq 1/\sqrt{2},$ we have $1 / \sqrt{2} \leq a \leq 1$ and $0 \leq m \leq 1.$
A direct evaluation gives
\beq
\begin{aligned}
\mathcal{V} (l,\gamma) &= \frac{L^2 a}{2} \int_1^{\bar{\zeta}} 
\frac{ (\log \bar{\zeta} -\log \tau)  \, \tau^{1/2}}{\sqrt{(\tau-1)(\tau-m)} } \,  d \tau  \\
&= L^2 \le \frac{l}{\delta} + 
\eta(\gamma)
\log \le   \frac{l}{2 a \, \delta} \ri 
+  \chi (\gamma)  \ri \, ,
\label{volume-difetto-totale}
\end{aligned}
\eeq
where
\beq
\eta(\gamma)=a \left[ \int_1^{\infty} \tau^{1/2 } \le \frac{1}{\sqrt{(\tau-1) (\tau-m)}}-\frac{1}{\tau} \ri d \tau -2 \right]
=2 a \le \mathbb{K}(m)- \mathbb{E}(m)  \ri  \, ,
\label{eta-gamma-eq}
\eeq
\beq
\chi (\gamma)=a \left[ -2
 -\frac{1}{2} \int_1^{\infty} \tau^{1/2 } \log \tau \le \frac{1}{\sqrt{(\tau-1) (\tau-m)}}-\frac{1}{\tau} \ri d \tau  \right] \, .
\eeq
The  complexity of the total space can be obtained from the $l \rightarrow \infty$ limit of
Eq.~(\ref{volume-difetto-totale}).

To compute the contribution to the  volume arising purely
 from the defect, we need to subtract the result from $\mathrm{AdS}_3$ space.
This amounts to put $\gamma=0$ in the previous result, giving
\beq
\eta(0)=0 \, , \qquad \chi (0)= - \pi  \, .
\label{eq:volume_empty_AdS3_one_cutoff}
\eeq
For a cross-check, we can perform directly the AdS$_3$ calculation
\beq
y^* = \mathrm{arccosh} \, \le \frac{z}{\delta} \ri \, , \qquad
z_{\rm min} = \delta \, ,
\eeq
and the integral computing the volume is
\beq
\mathcal{V} (l,0) = 2 L^2 \int_{\delta}^{\bar{z}} \frac{dz}{z}
 \int_0^{\mathrm{arccosh (z/\delta)}} dy \, \cosh y  =
L^2\le \frac{l}{\delta} - \pi \ri \, ,
\label{V0-ads}
\eeq
which is consistent with Eq.~(\ref{eq:volume_empty_AdS3_one_cutoff}).

\begin{figure}
\begin{center}
 \def\svgwidth{\columnwidth}
    \scalebox{0.55}{
\begingroup%
  \makeatletter%
  \providecommand\color[2][]{%
    \errmessage{(Inkscape) Color is used for the text in Inkscape, but the package 'color.sty' is not loaded}%
    \renewcommand\color[2][]{}%
  }%
  \providecommand\transparent[1]{%
    \errmessage{(Inkscape) Transparency is used (non-zero) for the text in Inkscape, but the package 'transparent.sty' is not loaded}%
    \renewcommand\transparent[1]{}%
  }%
  \providecommand\rotatebox[2]{#2}%
  \newcommand*\fsize{\dimexpr\f@size pt\relax}%
  \newcommand*\lineheight[1]{\fontsize{\fsize}{#1\fsize}\selectfont}%
  \ifx\svgwidth\undefined%
    \setlength{\unitlength}{749bp}%
    \ifx\svgscale\undefined%
      \relax%
    \else%
      \setlength{\unitlength}{\unitlength * \real{\svgscale}}%
    \fi%
  \else%
    \setlength{\unitlength}{\svgwidth}%
  \fi%
  \global\let\svgwidth\undefined%
  \global\let\svgscale\undefined%
  \makeatother%
  \begin{picture}(1,0.65020027)%
    \lineheight{1}%
    \setlength\tabcolsep{0pt}%
    \put(0,0){\includegraphics[width=\unitlength,page=1]{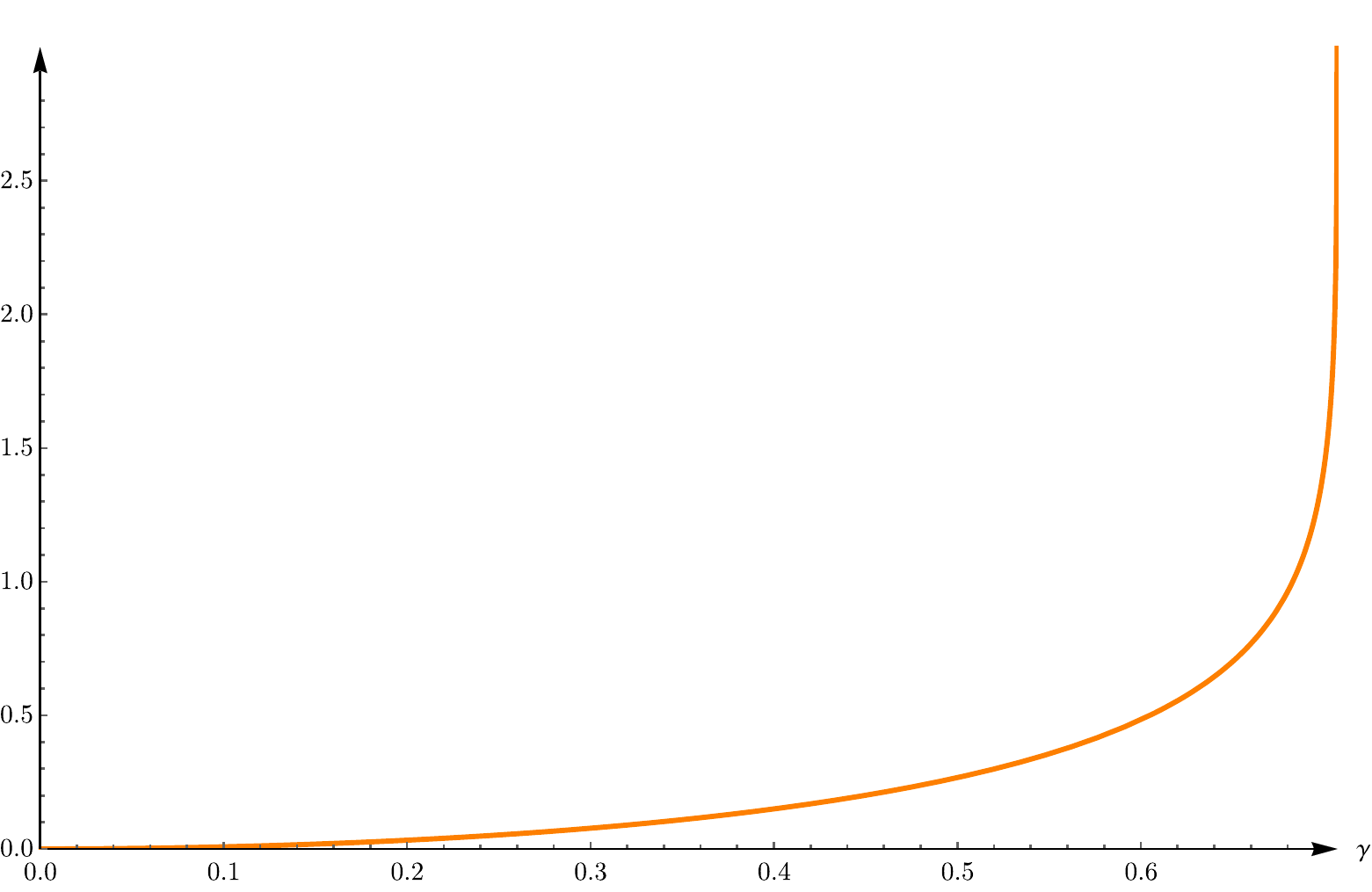}}%
    \put(0.01096723,0.62591568){\makebox(0,0)[lt]{\lineheight{1.25}\smash{\begin{tabular}[t]{l}$\eta(\gamma)$\end{tabular}}}}%
    \put(0,0){\includegraphics[width=\unitlength,page=2]{eta_di_gamma.pdf}}%
    \put(0.97765045,0.02271783){\makebox(0,0)[lt]{\lineheight{1.25}\smash{\begin{tabular}[t]{l}$\gamma$\end{tabular}}}}%
  \end{picture}%
\endgroup%
}
\caption{
Plot of $\eta(\gamma)$ as defined in Eq.~(\ref{eta-gamma-eq}), which is the coefficient
of the log divergencies due to the defect. 
}
\label{eta-gamma}
\end{center}
\end{figure}

The difference between the regularised volumes of the Janus geometry and AdS$_3$ is
\beq
\Delta \mathcal{V}(l,\gamma) \equiv \mathcal{V}(l,\gamma) - \mathcal{V}(l,0) =
L^2 \le \eta(\gamma) \,
\log \le    \frac{l}{2 a \,  \delta} \ri
 + \chi (\gamma) + \pi   \ri \, .   
 \label{eq:deltaV_one_cutoff}
\eeq
The only  divergence in Eq.~(\ref{eq:deltaV_one_cutoff})  is logarithmic, 
which is interpreted as the contribution from the defect,
and it is proportional to the function $\eta(\gamma)$.
Note that for small $\gamma$
\beq
\eta(\gamma) \approx \frac{\pi}{4} \gamma^2 \, ,
\eeq
and that $\eta$ is divergent for  $\gamma \rightarrow 1/\sqrt{2}$, which corresponds to the 
linear dilaton limit.
A plot of $\eta(\gamma)$ is shown in figure \ref{eta-gamma}.

Note that the $l$-dependent part of Eq.~(\ref{eq:deltaV_one_cutoff})
is proportional to the entanglement entropy of the segment without the defect, i.e.,
\beq
S_{\mathrm{AdS}} = \frac{c}{3} \log \le \frac{l}{\delta} \ri \, .
\eeq

\subsection{Fefferman-Graham regularization}
\label{app-static_volume_FG_method}

This method consists in using the FG form of the metric in eq. \eqref{eq:form_metric_FGexpansion} 
to identify a UV cutoff in terms of the proper radial coordinate near the boundary.
In particular, it is required that when $\xi \rightarrow 0$ the asymptotic behaviour respects the limits
\beq
g_1 ( \xi/\eta ) \rightarrow 1 \, , \qquad
g_2 ( \xi/\eta ) \rightarrow 1 \, .
\eeq
The change of coordinates from $(y,z)$ to $(\xi, \eta)$ breaks down when $ \xi/\eta \gg 1,$ which corresponds to approaching the interface.
This condition is equivalent to the statement that there exists a value of $y=y_0$ such that the FG expansion breaks down.
Since $f(y)$ is a monotonically increasing function in the region $y \geq 0,$ this equivalently implies that there exists a value  $b \geq 1$ such that
\beq
A(y_0) = b \, A(0)  \, ,
\eeq
where $A(y)$ was introduced in Eq.~\eqref{eq:metric_Trivella_form}.
This criterion selects a particular $y=y_0$ such that
\beq
y_0 = A^{-1} ( b A(0)) \, .
\label{eq:definition_y0_FG_patches}
\eeq
There exist universal quantities that do not depend on the choice of the curve connecting the two FG patches \cite{Estes:2014hka}.
For this reason, in the region where $y \in [-y_0, y_0]$ we can introduce an arbitrary curve interpolating between the two regions.
As proposed in \cite{Chapman:2018bqj}, we will select an interpolating curve connecting smoothly the two patches.
In the following, we show that this prescription gives the same result as the single cutoff 
method applied in Section \ref{sect-Janus_one_cutoff}, except for  the finite part.

\subsubsection*{Integration in the FG patches}

We consider the FG expansion defined in Eq.~\eqref{eq:change_coordinates_FG_2} with a UV cutoff located at $\xi= \delta$ and the condition $y \gg 1.$
In this way we find
\beq
\frac{\delta}{z}  =    \frac{1}{\sqrt{f(y)}} \, , \quad \Rightarrow 
 \quad y^* = \frac{1}{2} \, \mathrm{arccosh}  \le \frac{ \frac{2 z^2}{\delta^2} -1}{\sqrt{1-2\gamma^2}}\ri   \, ,
\eeq
which is equivalent to the single cutoff prescription \eqref{eq:single_cutoff_prescription} after using the 
identification in \eqref{eq:dictionary_Janus_Trivella}.
According to Eq.~\eqref{eq:definition_y0_FG_patches}, we determine the minimal value of $y=y_0$ such that the FG expansion is valid by solving
\beq
\sqrt{f(y_0)} = a \, b\quad  \Rightarrow \quad
y_0 = \frac{1}{2} \, \mathrm{arccosh} \, \le \frac{2 a^2 b^2 -1}{\sqrt{1-2 \gamma^2}}\ri \, .
\label{eq:condition_y0_Janus}
\eeq
Therefore, the integration in the FG patch region is given by
\beq
\mathcal{V}_{\rm FG}^1 (l,\gamma) = 2 L^2 \int_{z^{\rm FG}_{\rm min}}^{\bar{z}} \frac{dz}{z} \int_{y_0}^{y^*(z)} dy \, \sqrt{f(y)} \, ,
\eeq
where $\bar{z}$ was defined in Eq.~\eqref{eq:definition_zbar}, and  we put a symmetry factor of 2 due to the symmetry of the problem.
The minimal value $z=z^{\rm FG}_{\rm min}$ is determined as
\beq
\label{cutoffz}
z_{\rm min}^{\rm FG} = \underset{y \in [y_0, y^*]}{\mathrm{min}} \, \le \sqrt{f(y)} \ri  \delta 
 = a \, b \, \delta   \, ,
\eeq
where we used the fact that in a single FG patch the function $f(y)$ is monotonically increasing.
Notice that this value of $z^{\rm FG}_{\rm min}$ differs from $z_{\rm min}$ in Eq.~\eqref{eq:ystar_Janus_one_cutoff} (determined for the single cutoff prescription) only by the factor $b.$

\subsubsection*{Interpolation in the middle region}

Now we consider the middle region where we do not have access to a FG expansion.
We show that the surfaces at constant $y$ and the ones at constant $z$ are orthogonal to each others.
Since the normal one-forms to such surfaces are given by
\beq
\mathbf{v} = dy \, , \qquad
\mathbf{w} = dz \, ,
\eeq 
one can easily show that $\mathbf{v} \cdot \mathbf{w}=0.$ 
Even though the coordinates $(\xi, \eta)$ are not defined in the middle region, the original variables $(y,z)$ are still valid.
According to \cite{Chapman:2018bqj}, the curve interpolating the FG patches should be chosen perpendicular to the surface located  at $y=y_0.$
On the time slice $t=0,$ this condition selects curves at constant $z.$

The integral in this region reads
\beq
\mathcal{V}^2_{\rm FG}(l,\gamma) = 2  L^2 \int_{z^{\rm FG}_{\rm min}}^{\bar{z}} \frac{dz}{z} \int_0^{y_0} dy \, \sqrt{f(y)} \, .
\eeq
The total volume is
\beq
\label{FGreg}
\mathcal{V}_{\rm FG} (l,\gamma) \equiv \mathcal{V}^1_{\rm FG} (l,\gamma) + \mathcal{V}^2_{\rm FG} (l,\gamma) =
2 L^2 \int_{z^{\rm FG}_{\rm min}}^{\bar{z}} \frac{dz}{z} \int_0^{y^*} dy \, \sqrt{f(y)} \, ,
\eeq
which is exactly the integral \eqref{eq:integral_one_cutoff} that we evaluated for the single cutoff prescription, except that now we integrate from $z_{\rm min}^{\rm FG} \geq z_{\rm min}. $ 
Subtracting vacuum $\mathrm{AdS}_3,$ the result is
\beq
\Delta \mathcal{V}_{\rm FG} (l,\gamma)=L^2 \left[
\eta (\gamma) \log \le  \, \frac{l}{2  a \, b \, \delta} \ri  
 + \chi (\gamma) + \pi  \right] \, .   
 \label{eq:deltaV_FG}
\eeq
The result, up to a finite term, coincides with Eq.~(\ref{eq:deltaV_one_cutoff}).
This procedure also shows that the value $y=y_0$ where the FG patch ends does not play any special role.

\subsection{Double cutoff regularization}
\label{app-Janus_AdS3_double_cutoff}
Here we repeat the calculation of the previous section
 using the double cutoff prescription.
Following the procedure described in Section \ref{sect-different_regularizations},
 we take one cutoff at $z= \delta$, whereas the other one is determined by
\beq
f(y) = \frac{1}{\varepsilon^2} \, .
\label{effe-epsilon}
\eeq
Solving the previous equations for $y=y^*$, we find
\beq
y^*(\varepsilon) =f^{-1}\left( \frac{1}{\varepsilon^2}\right) \, .
\eeq
This value is the same as the one in Eq.~\eqref{eq:ystar_Janus_one_cutoff}, once we identify $\varepsilon = \delta / z .$
The main difference using this method is that $\varepsilon$ does not depend on $z$. Thus, the two integrals defining the volume are independent and factorize.
Then, the extremal volume for the subregion is given by 
\beq
\mathcal{V} (l,\gamma) = 2 L^2 \int_{\delta}^{\bar{z}} \frac{dz}{z} \int_0^{y^*(\varepsilon)} dy \, \sqrt{f(y)} \, .
\eeq
 Following the analysis in Section \ref{sect-Janus_one_cutoff} and performing the change of variables
\beq
\tau=\frac{f(y)}{a^2}\;,
\eeq
we obtain
\beq
\mathcal{V}(l,\gamma)=L^2 a \log\left(\frac{\bar{z}}{\delta} \right) \int_{1}^{{1}/{(a^2\varepsilon^2)}} d\tau \sqrt{\frac{\tau}{(\tau-1)(\tau-m)}}\;.
\eeq
A direct evaluation gives
\beq
\mathcal{V}(l,\gamma)=L^2\left(\eta(\gamma) +\frac{2}{\varepsilon}\right)  \log\left(\frac{\bar{z}}{\delta} \right)\;,
\eeq
where $\eta(\gamma)$ is defined in Eq.~\eqref{eta-gamma-eq}. Subtracting the volume of pure AdS$_3$ space which is given by
\beq
\mathcal{V}(l,0)= L^2\frac{2}{\varepsilon}  \,  \log \le \frac{\bar{z}}{\delta} \ri + \mathcal{O}(\varepsilon) \,,
\eeq
we get the contribution to the volume from the defect
\beq
\Delta \mathcal{V}(l,\gamma)= L^2\, \eta(\gamma)\log\left(\frac{l}{2\,\delta} \right)+\mathcal{O}(\delta)\,.
\eeq
We observe that the dependence on $\varepsilon$ disappears after the subtraction, consistently with the non-physical relevance of this cutoff \cite{Gutperle:2016gfe}. 
The remaining logarithmic divergence matches with the  single cutoff calculation
 in Eq.~\eqref{eq:deltaV_one_cutoff},
up to finite parts.

\subsection{Comparison with other defect geometries}
\label{comparison-other-geometries}
In this section, we investigated subregion volume complexity 
for an interval of length $l$ centered around the Janus interface, using three different regularizations.
We found that the increment of subregion complexity compared to the vacuum CFT is
\beq
\Delta \mathcal{C}(l,\gamma)  =
\frac{2}{3} c \,  \eta(\gamma) \,
\log \le    \frac{l}{\,  \delta} \ri
 + \text{finite terms}  \, ,
 \label{result-janus-Tzero}
\eeq
where $c$ is the CFT central charge and $\eta(\gamma)$
is
\beq
\eta(\gamma) =2 
\sqrt{ \frac{1+\sqrt{1-2\gamma^2}}{2 }}
\,  \left[ \mathbb{K}(m)- \mathbb{E}(m)  \right]  \, ,
 \qquad
m = \frac{1 - \sqrt{1-2\gamma^2}}{1 + \sqrt{1-2\gamma^2}} \, .
\eeq
A plot of $\eta(\gamma)$ is shown in figure
\ref{eta-gamma}.
We can contrast this with the ground state degeneracy $g$ of the Janus solution computed in \cite{Azeyanagi:2007qj}, given by
\beq
\Delta S =\log g=\frac{c}{6} \kappa(\gamma) \, , \qquad \kappa(\gamma)=\log \frac{1}{\sqrt{1-2 \gamma^2}} \, .
\label{g-difetto-kappa}
\eeq
The value $0 \leq \gamma < \frac{\sqrt{2}}{2}$  parameterizes the excursion
of the dilaton between the two sides of the interface,
which diverges for $\gamma \rightarrow \frac{\sqrt{2}}{2} $.

\begin{figure}[h]
\begin{center}
\def\svgwidth{\columnwidth}
    \scalebox{0.6}{
\begingroup%
  \makeatletter%
  \providecommand\color[2][]{%
    \errmessage{(Inkscape) Color is used for the text in Inkscape, but the package 'color.sty' is not loaded}%
    \renewcommand\color[2][]{}%
  }%
  \providecommand\transparent[1]{%
    \errmessage{(Inkscape) Transparency is used (non-zero) for the text in Inkscape, but the package 'transparent.sty' is not loaded}%
    \renewcommand\transparent[1]{}%
  }%
  \providecommand\rotatebox[2]{#2}%
  \newcommand*\fsize{\dimexpr\f@size pt\relax}%
  \newcommand*\lineheight[1]{\fontsize{\fsize}{#1\fsize}\selectfont}%
  \ifx\svgwidth\undefined%
    \setlength{\unitlength}{634bp}%
    \ifx\svgscale\undefined%
      \relax%
    \else%
      \setlength{\unitlength}{\unitlength * \real{\svgscale}}%
    \fi%
  \else%
    \setlength{\unitlength}{\svgwidth}%
  \fi%
  \global\let\svgwidth\undefined%
  \global\let\svgscale\undefined%
  \makeatother%
  \begin{picture}(1,0.65299685)%
    \lineheight{1}%
    \setlength\tabcolsep{0pt}%
    \put(0,0){\includegraphics[width=\unitlength,page=1]{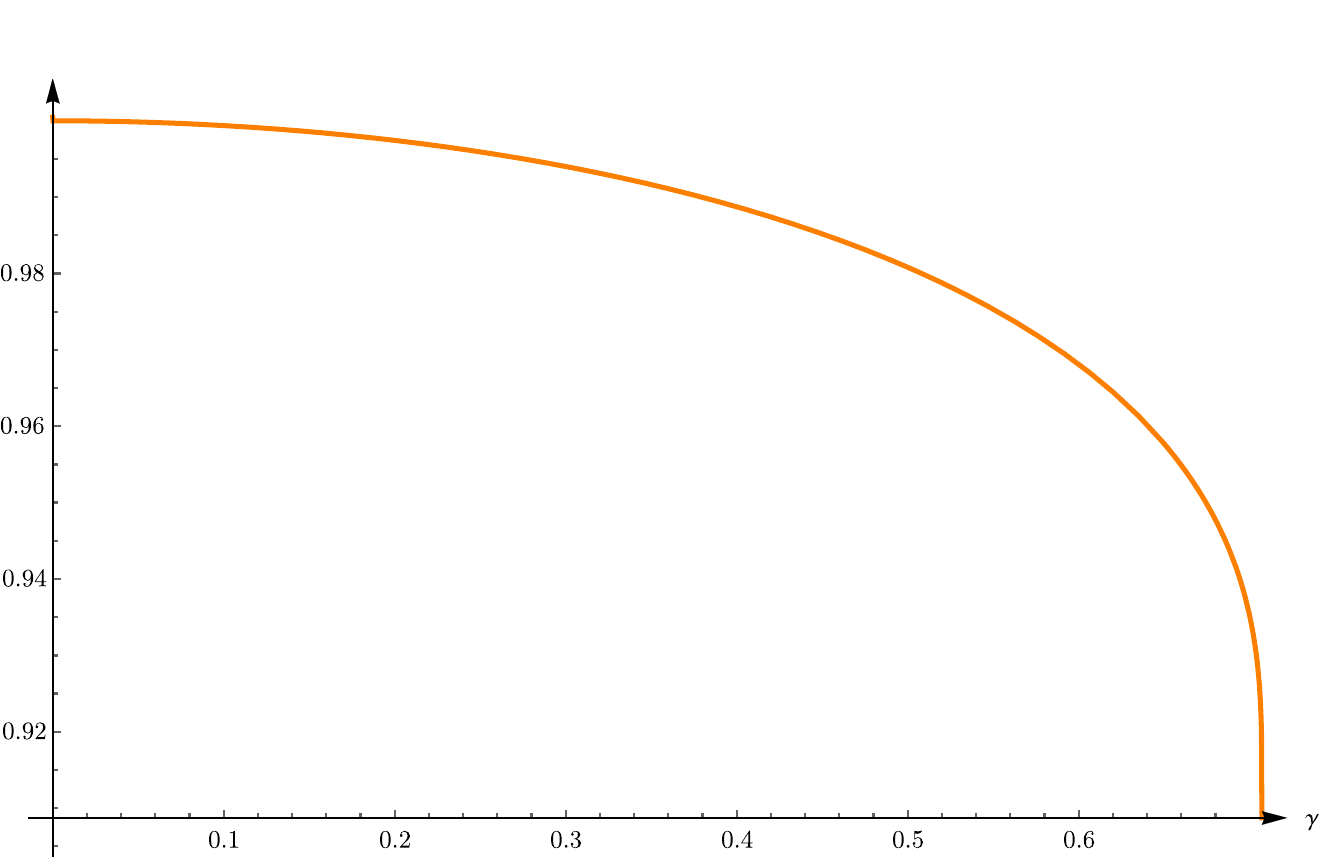}}%
    \put(0.01780197,0.6118298){\makebox(0,0)[lt]{\lineheight{1.25}\smash{\begin{tabular}[t]{l}$\frac{4\eta}{\pi\kappa}$\end{tabular}}}}%
    \put(0,0){\includegraphics[width=\unitlength,page=2]{eta_su_kappa.pdf}}%
    \put(0.97333658,0.02378095){\makebox(0,0)[lt]{\lineheight{1.25}\smash{\begin{tabular}[t]{l}$\gamma$\end{tabular}}}}%
  \end{picture}%
\endgroup%
}
\caption{
Plot of $4 \eta/(\pi \kappa)$ as a function of $\gamma$,
see Eq.~(\ref{g-difetto-kappa}).
}
\label{eta-gamma-2}
\end{center}
\end{figure}

It is interesting to compare $\Delta \mathcal{C}$ for the Janus interface with recent results found for other defect geometries, namely:
\begin{itemize}
\item the AdS$_3$ Randall-Sundrum  model, where the contribution to complexity coming from the defect, evaluated in \cite{Chapman:2018bqj}, is
\beq
\Delta \mathcal{C}(l, y^* ) = \frac{2}{3} c \, \eta_{_\text{RS}} \, \log \frac{l}{\delta} \,  + \text{finite terms} \, ,
\qquad
\eta_{_\text{RS}} =2 \sinh y^* \, ,
\eeq
where the parameter $0 \leq y^* < \infty $ is related to the brane tension $\l$ by
\beq
\l=\frac{\tanh y^*}{ 4 \pi G L } \, .
\eeq
In this case, the defect boundary entropy \cite{Azeyanagi:2007qj} is
\beq
\Delta S = \log g = \frac{c}{6}  \kappa_{_\text{RS}} \, , \qquad \kappa_{_\text{RS}}=2 y^* \, ;
\eeq

\item the BCFT model. Complexity was studied in \cite{Sato:2019kik, Braccia:2019xxi} and a similar scenario for an holographic Kondo model was investigated in \cite{Flory:2017ftd}.
For the three-dimensional case, the subregion volume complexity reads
\beq
\Delta \mathcal{C}(l, \a ) = \frac{2}{3} c \, \eta_{_\text{BCFT}} \, \log \frac{l}{\delta} \,  + \text{finite terms} \, ,
\qquad
\eta_{_\text{BCFT}}= \cot \a \, ,
\eeq
where the parameter $ 0 < \alpha \leq   \frac{\pi}{2 } $ is related to the brane tension $T$ as follows
\beq
T=\frac{\cos \a}{ L} \, .
\eeq
The defect boundary entropy  \cite{Takayanagi:2011zk,Fujita:2011fp} is
\beq
\Delta S = \log g = \frac{c}{6} \kappa_{_\text{BCFT}} \, ,
\qquad
\kappa_{_\text{BCFT}} =  \log \le \cot \frac{\a}{2}\ri  \, .
\eeq
\end{itemize}
We conclude that:
\begin{enumerate}
\item the leading divergence in  $\Delta \mathcal{C}$  is always logarithmic,  with a positive coefficient $\eta$
that is a function of the deformation parameter of the model;
\item in all three cases, the log divergence in $\Delta \mathcal{C}$
is not related in any universal way to the defect boundary entropy $\Delta S$.
Nonetheless, the two quantities share a similar behavior, that is, they diverge for the same critical value;
\item for small values of the deformation parameters, in all the models $\eta/\kappa$ is of order $1$, i.e.,
\beq
\frac{\eta(\gamma)}{\kappa(\gamma)}=\frac{\pi}{4} \, , \qquad
\frac{\eta_{_\text{RS}}}{\kappa_{_\text{RS}}}=\frac{\eta_{_\text{BCFT}}}{\kappa_{_\text{BCFT}}}=1 \, ;
\eeq
\item in the Janus case $\eta/\kappa$ remains very close to $\frac{\pi}{4}$ for the whole range
of the deformation parameter $\gamma$ (see figure \ref{eta-gamma-2}). On the contrary, in the other two models $\eta \gg \kappa$ close to the critical values $y^* \to \infty$ and $\a \to 0$.
\end{enumerate}


\section{Volume of the static Janus BTZ black hole}
\label{section-static-BTZ}

In the finite temperature case, it is interesting to consider the thermal
complexity of formation \cite{Chapman:2016hwi}, which corresponds to the
additional complexity needed to prepare the thermofield double state 
starting from two copies of unentangled vacuum CFTs.
We study this quantity for the Janus interface.
In the CV conjecture, this corresponds to the volume of the initial time slice
of the Kruskal extension of the static Janus BTZ subtracted by 
twice the Janus AdS contribution.

\subsection{Single cutoff regularisation}
\label{sect-volume_static_JBTZ}
In this Section we compute the subregion complexity
for a symmetric interval with length $l$ centered around the defect
 of the static Janus BTZ geometry at vanishing boundary time. 
 The  metric is
\begin{align}
ds^2= L^2 \left[dy^2+f(y)\left(-\frac{(w^2-1)r_h^2}{L^4}dt^2+\frac{dw^2}{w^2-1}\right)\right],
\end{align}
which is obtained by substituting the Rindler-like metric \eqref{eq:Rindler_slicing} 
in \eqref{eq:general_form_3d_Janus}, and performing the $\mu$--to--$y$ change of variable $d\mu=dy/\sqrt{f(y)}$.

The HRT surface on a constant time slice lies at a constant value $w=\bar{w}$ 
\cite{Bak:2020enw}.
We locate the interval along the coordinate $x\in[-\frac{l}{2},\frac{l}{2}]$ orthogonal to the defect and 
on the real boundary parametrized by $r=\infty.$
The value of $\bar{w}$ can be obtained by combining the two equations  \eqref{eq:asymp_JBTZ}
with $r=\infty$, $x=l/2$ and solving for $\bar{w}$, which gives
\begin{align}
\bar{w}=\coth \left(\frac{l r_h}{2 L^2}\right)
\label{eq:definition_wbar_JBTZ}
\end{align}
The UV divergencies are regularized 
with the single cutoff prescription presented in Section \ref{sect-different_regularizations}. 
The natural choice of cutoff is found by performing a FG expansion of the metric to relate the asymptotic behavior of the deformed BTZ black hole with the non-deformed counterpart.
Such asymptotic behaviour is identified by 
\begin{align}
\frac{r}{r_h} \simeq \sqrt{(w^2-1)f(y)+1} \, . 
\end{align}
 The cutoff surface determined by the FG expansion at $r=L^2/\delta$ 
 induces the following value of $y$ coordinate
\begin{align}
\label{eq:1cutoff_JBTZ}
y^*(w)=\frac{1}{2}\text{arcosh}\left(\frac{2-\hat{\delta}^2\left(w^2+1\right)}{\hat{\delta}^2 \left(w^2-1\right) \,  \sqrt{1-2 \gamma ^2} }\right) \, , \qquad
\hat{\delta}=\frac{r_h \delta }{L^2} \, .
\end{align}
This in turn induces a cutoff in the $w$ coordinate, which is found by maximizing the inverse of the previous function
\begin{align}
w(y)=\sqrt{\frac{{2}/{\hat{\delta}^2 }+\sqrt{1-2 \gamma ^2} \cosh (2 y)-1}{\sqrt{1-2 \gamma ^2} \cosh (2 y)+1}},
\end{align}
with respect to $y$. The maximum occurs at $y=0$, and thus 
\begin{align}
\label{eq:max_JBTZ}
w_\text{max}=\sqrt{\frac{{2}/{\hat{\delta}^2 }+\sqrt{1-2 \gamma ^2}-1}{\sqrt{1-2 \gamma ^2}+1}}\, .
\end{align} 
This UV cutoff is the analog of $z_{\rm min}$ in the Janus AdS$_3$ case, see Eq.~\eqref{eq:ystar_Janus_one_cutoff}.

The extremal volume \eqref{eq:general_prescription_volume} for the static Janus BTZ geometry reads
\begin{align}
\mathcal{V} (l,\gamma)=2 L^2\int_{\bar{w}}^{w_\text{max}} d w\int_{0}^{y^*(w)}dy \,\sqrt{\frac{f(y)}{w^2-1}} \, .
\end{align}
After performing the following change of variables
\beq
\tau = \frac{f(y)}{a^2} \, , \qquad
z = \frac{1-\hat{\delta}^2}{a^2 \hat{\delta}^2  (w^2-1)} \, , \qquad
\bar{z} = \frac{1-\hat{\delta}^2}{a^2 \hat{\delta}^2  (\bar{w}^2-1)}  \, ,
\eeq
the integral takes the form
\beq
\begin{aligned}
\mathcal{V} (l,\gamma) & = \frac{L^2 a}{2} \int_1^{\bar{z}} \frac{dz}{\sqrt{1+ \alpha^2 z}} 
\int_1^{z} d\tau \, \sqrt{\frac{\tau}{(\tau-1)(\tau-m)}}  \\
& =  \frac{L^2 a}{2} \int_1^{\bar{z}} d\tau \sqrt{\frac{\tau}{(\tau-1)(\tau-m)}} \int_{\tau}^{\bar{z}} \frac{dz}{z \sqrt{1+\alpha^2 z}}  \, ,
\end{aligned}
\eeq
where 
\beq
\alpha = \frac{a  \, \hat{\delta}}{\sqrt{1-\hat{\delta}^2 }} \, .
\eeq
The integration over $z$ yields three kind of terms, according to which we split the extremal volume as
\beq
\mathcal{V} (l,\gamma) = \mathcal{V}_1 + \mathcal{V}_2 + \mathcal{V}_3 \, ,
\eeq
where
\beq
\begin{aligned}
\mathcal{V}_1 &= L^2 a \int_1^{\bar{z}} d\tau \,  \log \le 1+\sqrt{1+\alpha^2 \tau} \ri \sqrt{\frac{\tau}{(\tau-1)(\tau-m)}} \, ,
\\
\mathcal{V}_2& = - \frac{L^2 a}{2} \int_1^{\bar{z}} d\tau \,  \log \tau  \,  \sqrt{\frac{\tau}{(\tau-1)(\tau-m)}} \, ,
\\
\mathcal{V}_3 &=  -L^2 a \int_1^{\bar{z}} d\tau \, \log \le \frac{1+\sqrt{1+\alpha^2 \bar{z}}}{\sqrt{\bar{z}}} \ri
 \sqrt{\frac{\tau}{(\tau-1)(\tau-m)}}  \, .
\end{aligned}
\eeq
These integrals can be evaluated explicitly following similar steps as in 
section \ref{sect-Janus_one_cutoff}. The result is
\beq
\mathcal{V} (l,\gamma) = L^2 \le  \, \frac{l}{\delta}
 + \eta(\gamma) \log \left[ \frac{2L^2}{a \, \delta \,  r_h} \tanh \le \frac{l r_h}{4L^2} \ri \right]
+  \chi (\gamma) 	\ri + \mathcal{O} (\delta) \, ,
\eeq 
where $\eta(\gamma)$ and $\chi (\gamma)$ were defined in Eq.~\eqref{eta-gamma-eq}.

In absence of the interface (for $\gamma=0$)
the volume  is still given
by the AdS$_3$ result in Eq.~(\ref{V0-ads}). This is due to the fact that subregion volume complexity in the 
BTZ background does not depend on temperature and it is topologically protected
by the Gauss-Bonnet theorem \cite{Abt:2017pmf}. On the other hand, in the case of the Janus interface subregion complexity depends on the temperature.
In the small temperature regime $r_h l \ll 1,$ we recover the Janus AdS
 result in Eq.~\eqref{volume-difetto-totale}. 

\begin{figure}
\begin{center}
 \def\svgwidth{\columnwidth}
    \scalebox{0.55}{
\begingroup%
  \makeatletter%
  \providecommand\color[2][]{%
    \errmessage{(Inkscape) Color is used for the text in Inkscape, but the package 'color.sty' is not loaded}%
    \renewcommand\color[2][]{}%
  }%
  \providecommand\transparent[1]{%
    \errmessage{(Inkscape) Transparency is used (non-zero) for the text in Inkscape, but the package 'transparent.sty' is not loaded}%
    \renewcommand\transparent[1]{}%
  }%
  \providecommand\rotatebox[2]{#2}%
  \newcommand*\fsize{\dimexpr\f@size pt\relax}%
  \newcommand*\lineheight[1]{\fontsize{\fsize}{#1\fsize}\selectfont}%
  \ifx\svgwidth\undefined%
    \setlength{\unitlength}{634bp}%
    \ifx\svgscale\undefined%
      \relax%
    \else%
      \setlength{\unitlength}{\unitlength * \real{\svgscale}}%
    \fi%
  \else%
    \setlength{\unitlength}{\svgwidth}%
  \fi%
  \global\let\svgwidth\undefined%
  \global\let\svgscale\undefined%
  \makeatother%
  \begin{picture}(1,0.62460568)%
    \lineheight{1}%
    \setlength\tabcolsep{0pt}%
    \put(0,0){\includegraphics[width=\unitlength,page=1]{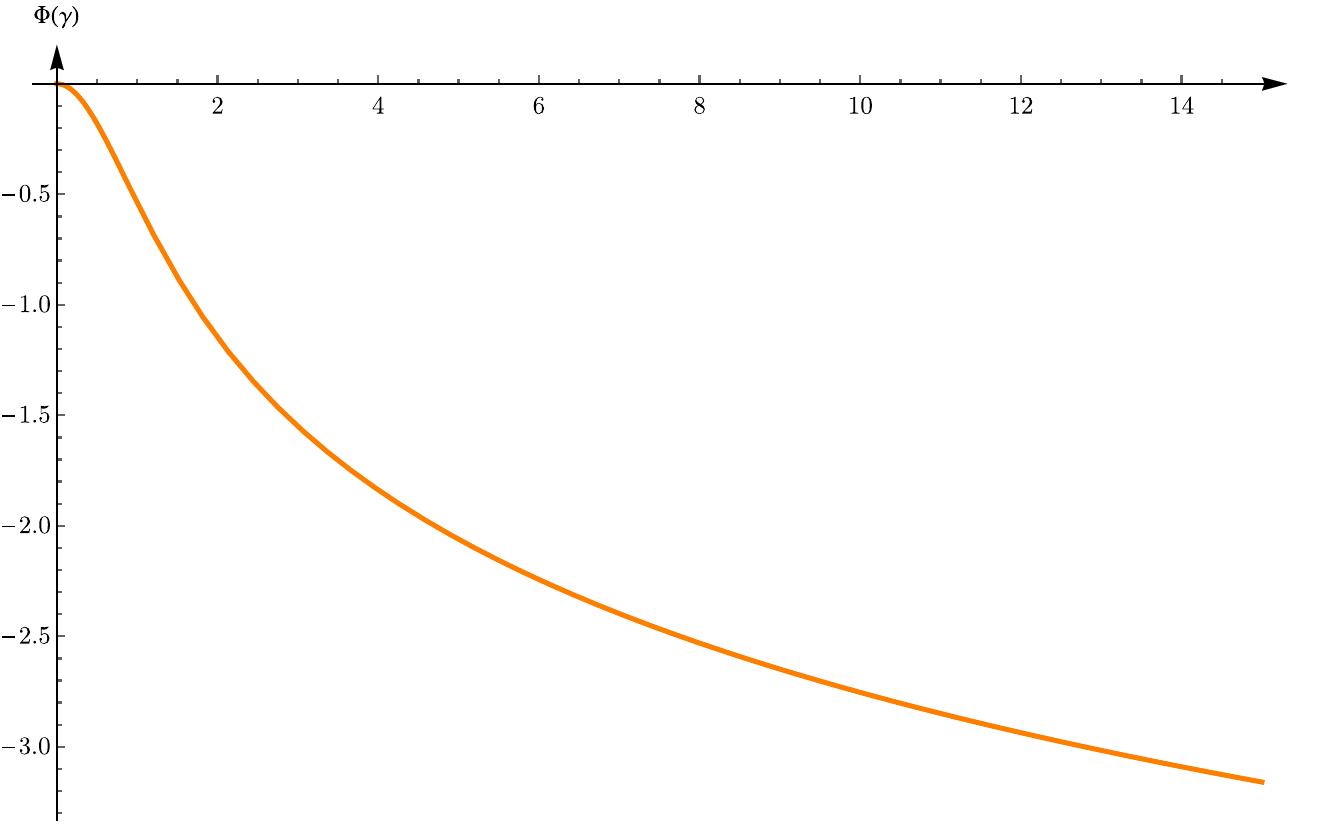}}%
    \put(0.97679286,0.55160959){\makebox(0,0)[lt]{\lineheight{1.25}\smash{\begin{tabular}[t]{l}$Tl$\end{tabular}}}}%
    \put(0,0){\includegraphics[width=\unitlength,page=2]{phi_di_gamma.pdf}}%
    \put(0.02332865,0.59751435){\makebox(0,0)[lt]{\lineheight{1.25}\smash{\begin{tabular}[t]{l}$\Phi(Tl)$\end{tabular}}}}%
  \end{picture}%
\endgroup%
}
\caption{
Plot of $\F(T l)$ in Eq.~(\ref{tempura2}), which gives the temperature
dependence of the complexity of the interface.
}
\label{comple-temperatura}
\end{center}
\end{figure}

Using the expression for  temperature and the central charge 
\beq
T=\frac{r_h}{2 \pi L^2} \, , \qquad c=\frac{3 L}{2 G} \, ,
\eeq
we get that the subregion complexity, in terms of field theory quantities, is
\begin{align}
\mathcal{C} (T, l,\gamma) =\frac{\mathcal{V} (l,\gamma) }{L G}=
 \frac23 c \le
 \frac{ l }{\delta} + \eta(\gamma)
  \log \left[ \frac{1}{\pi a T  \delta} \tanh \left(\frac{\pi l T}{2 } \right) \right] + \chi (\gamma) 
 \ri \, .
 \label{comple-BTZ-janus}
\end{align}
The temperature dependence of the volume complexity is
\beq
\Delta \mathcal{C} (T,l,\gamma) \equiv
\mathcal{C} (T,l,\gamma) -
 \mathcal{C} (0,l,\gamma) =
\frac23 \, c   \, \eta(\gamma)  \F(T l ) \,
 \label{tempura}
\eeq
where
\beq
\F(T l ) = \log \left[ \frac{2}{\pi   l T} \tanh \left(\frac{\pi l T}{2 } \right) \right] \, .
 \label{tempura2}
\eeq
At zero temperature, the contribution 
of the defect to $\mathcal{C} (0,l,\gamma) $ is proportional to the entanglement 
entropy of the segment without defect.
However, the proportionality is spoiled at finite  temperature.
In fact, the BTZ entanglement entropy for a segment of length $l$ is
\beq
S_{\rm BTZ}(T,l) = \frac{c}{3} \log \left[ \frac{2L^2}{r_h \, \delta} \sinh \le \frac{l r_h}{2L^2} \ri \right]
= \frac{c}{3}
\log \left[ \frac{1}{\pi  \, T \, \delta} \sinh \le \pi T l \ri \right] \, ,
\eeq
which is an increasing function of $T$, while $\Delta \mathcal{C} (l,\gamma) $ 
is a decreasing function of $T$ (see figure \ref{comple-temperatura}).

Taking into account the two sides of the Kruskal diagram,
the complexity of formation $\mathcal{C}_F$ of the static Janus BTZ starting from the Janus $\mathrm{AdS}_3$ 
background is given by the $l \gg 1/T$ limit of $2 \Delta \mathcal{C}$, that is, 
\beq
\mathcal{C}^{F}_{\mathrm{Thermal}}=
\frac{4}{3} c \,\eta(\gamma) \log \left( \frac{2}{\pi l T} \right) \, ,
\eeq
where, again, we interpret $l$ as an infrared regulator. 
In this limit, the complexity of formation is negative.
This means that it takes less complexity to form a BTZ black hole
 in the Janus geometry than the defect alone, starting from  empty space. 

It is also interesting to consider the complexity of formation of the defect in the BTZ background. 
In this case, we subtract from Eq.~\eqref{comple-BTZ-janus}
the $\gamma=0$ result
\beq
\mathcal{C} (T,l,\gamma) -
\mathcal{C} (T,l,0) =
\frac23 \, c   \,\le  \eta(\gamma)
\log \left[ \frac{1}{\pi a T  \delta} \tanh \left(\frac{\pi l T}{2 } \right) \right] + \chi (\gamma)+\pi \ri\,.
\label{eq:complexity_formation_with_gamma0_JBTZ}
\eeq
Then, considering the $lT\gg 1$ limit and multiplying by an additional factor of two to account for the two sides of the Kruskal diagram, we obtain the complexity of formation of the defect starting from the static BTZ background 
\beq
\mathcal{C}^{F}_{\mathrm{Defect}}  = 
\frac{4}{3} c \le \eta(\gamma) \log \left( \frac{1}{\pi a T \delta} \right)+ \chi (\gamma)+\pi  \ri\,.
\label{eq:complexity_formation_defect_section4_1}
\eeq
Notice that, since the volume complexity in the BTZ black hole is topological, the above result can be also interpreted as the complexity of formation of the Janus static BTZ starting from the vacuum $\mathrm{AdS}_3$ spacetime.

\subsection{Double cutoff regularisation}

We compute the subregion volume complexity for the static Janus BTZ geometry by employing the double cutoff regularization.
According to this prescription, we directly introduce a physical UV cutoff along the $w$ direction located at $w_{\rm max}= 1/\hat{\delta},$ but in addition we regularize the integration along $y$ by requiring eq. (\ref{effe-epsilon}).
This constraint induces a maximal value of $y=y^*$ that reads
\beq
y^* = \frac{1}{2} \mathrm{arccosh} \le \frac{2-\varepsilon^2}{\sqrt{1-2\gamma^2} \, \varepsilon} \ri \, .
\eeq
Finally, the integration is restricted by the HRT surface, that is located at the constant value $w=\bar{w}$ as determined in Eq.~\eqref{eq:definition_wbar_JBTZ}.
Collecting these geometrical data, we compute the volume as
\beq
\mathcal{V}(l,\gamma) = 2 L^2 \int_{\bar{w}}^{w_{\rm max}} dw \int_0^{y^*(\varepsilon)} \sqrt{\frac{f(y)}{w^2-1}} \, .
\eeq
Contrarily to the single cutoff case, the integrations are independent and can be evaluated separately.
Using the change of variables $\tau= f(y)/a^2,$ we find
\beq
\begin{aligned}
\mathcal{V}(l,\gamma) &= \frac{L^2 a}{2} \int_{\bar{w}}^{w_{\rm max}} \frac{dw}{\sqrt{w^2-1}}
\int_1^{1/(a^2 \varepsilon^2)} d\tau \, \sqrt{\frac{\tau}{(\tau-1)(\tau-m)}} = \\
& = L^2 \log \left[ \frac{1}{\pi T \delta} \tanh \le \frac{\pi l T}{2} \ri \right] 
\le \eta (\gamma) + \frac{2}{\varepsilon} \ri + \mathcal{O}(\delta, \varepsilon) \, .
\end{aligned}
\eeq
The complexity of formation compared to the BTZ black hole background is obtained by subtracting the result in the case $\gamma=0.$
It reads
\beq
\mathcal{C}(T,l,\gamma) - \mathcal{C}(T,l,0) = 
\frac{2}{3} c \le \eta(\gamma) \log \left[ \frac{1}{\pi T \delta} \tanh \le \frac{\pi l T}{2} \ri \right]  \ri + \mathcal{O}(\delta) \, ,
\eeq
whose divergence part precisely matches with the single cutoff computation, see Eq.~\eqref{eq:complexity_formation_with_gamma0_JBTZ}.
In addition, also the temperature dependence of the finite part is the same in both the regularizations:
their difference is  finite and 
depends only on $\gamma,$ as in eq.~\eqref{eq:complexity_formation_defect_section4_1}.

\section{Volume for the time-dependent Janus BTZ geometry}
\label{sect:time-dependent-BTZ}

In this section we will focus on the growth rate of complexity for the
 Janus BTZ geometry obtained through a time-dependent deformation of the BTZ black hole  \cite{Bak:2007jm,Bak:2007qw}.
 This gravity solution corresponds to a domain wall configuration for the dilaton field
along the radial direction of AdS, which connects the left and right sides
of the Penrose diagram. The dilaton does not divide each boundary component into two halves, rather, it takes two different values on each boundary.

The field theory dual of this solution is not an interface CFT,
but corresponds to two entangled CFTs
with two different but constant values of the dilaton source, 
$e^{\phi_-}$ and $e^{\phi_+}$, on each side
of the Penrose diagram.
 The two CFTs are correlated through the bulk in a non-trivial way
 and the solution  is time-dependent.
 In particular, the field theory dual starts at initial time
  from an out-of-equilibrium
 state, with an initial density for a non-conserved operator \cite{Bak:2007qw}.
 At later times, the theory thermalizes and approaches equilibrium.
The initial thermofield double state is determined by
\beq
| \psi (0,0) \rangle = \frac{1}{\sqrt{Z}} \sum_{m,n}
\langle E^L_m  | E_n^R \rangle \, 
e^{- \frac{\beta}{4} (E_m^L + E_n^R)} 
| E_m^L \rangle  \otimes | E_n^R \rangle \, .
\label{eq:TFD_Janus_BTZ_time_dependent}
\eeq
The time evolution of this state is not invariant under the boundary
time shift of Eq.~(\ref{bountary-time-shift}), because
the left and the right Hamiltonians are different.

The analysis is organized in the following way.
In section \ref{sect-BTZ_time_dep-preliminaries} we introduce the metric of the solution.
In Section \ref{sect-volume_vanishing_boundary_time_JBTZ_time_dep} we restrict 
to the $t=0$ case to analyze the structure of the UV divergences of the model.
Then, we move to the study of the time dependence of the volume by analysing the equations 
of motion and specifying the boundary conditions of the problem in Section \ref{sect-growth_rate_volume_analytic_part}.
 In Section \ref{sect-numerical_solutions_JBTZ_time_dep}
 we numerically determine the extremal slices and we plot the corresponding
  volume as a function of time.

\subsection{Preliminaries}
 \label{sect-BTZ_time_dep-preliminaries}

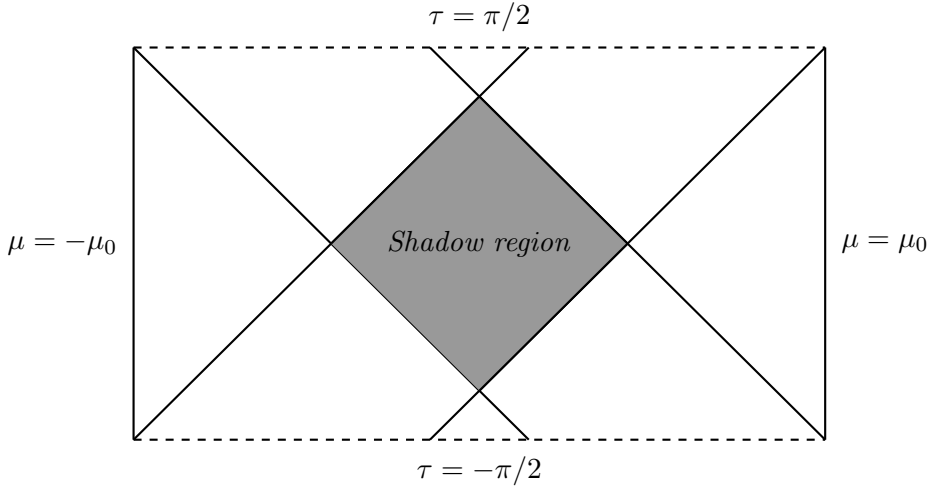
\begin{figure}[h]
\begin{center}
\begin{tikzpicture}[thick,scale=1.3]
\draw (-3.5,2) -- (-3.5,-2)
node[midway, left, inner sep=2mm] {$\mu=-\mu_0$};
\draw (-3.5,2) -- (0.5,-2);
\draw (0.5,2) -- (-3.5,-2);

\draw (3.5,2) -- (3.5,-2)
node[midway, right, inner sep=2mm] {$\mu=\mu_0$};
\draw (3.5,2) -- (-0.5,-2);
\draw (3.5,-2) -- (-0.5,2);
\fill[black!40!white,draw=black] (-1.5,0) -- (0,1.5) -- (1.5,0)--(0,-1.5);
\draw[black,thick,dashed] (-3.5,2) -- (3.5,2)
      node[midway, above, inner sep=2mm] {$\tau=\pi/2$};
\draw[black,thick,dashed] (-3.5,-2) -- (3.5,-2)
      node[midway, below, inner sep=2mm] {$\tau=-\pi/2$};
\node (I)    at ( 0,0)   {\textit{Shadow region}};
\end{tikzpicture}
\end{center}
\caption{Penrose diagram of the Janus deformation of the BTZ black hole.
 The $\mu$ variable runs along horizontal lines from $-\mu_0$ to $\mu_0$, while $\tau$ runs
  vertically from $-\pi/2$ to $\pi/2$. The shaded region represents the so-called \textit{shadow region}.}
\label{fig:Penrose_Janus_BTZ}
\end{figure}

In order to find the time-dependent background, let us start form the Janus AdS$_3$ solution in
Eq.~(\ref{eq:metric_Janus_mucoord})
 \beq
ds_3^2 = L^2 f(\mu)  \cos^2 \mu \, ds^2_{\mathrm{AdS}_3} \, ,
\qquad
ds^2_{\mathrm{AdS}_3} = \frac{1}{\cos^2 \mu} \le d\mu^2 + ds^2_{\mathrm{AdS}_2} \ri \, .
\eeq
We can replace the AdS$_3$ part of this geometry with the BTZ BH in Eq.~\eqref{eq:BTZ_metric}, 
as this solution is locally equivalent to pure AdS$_3$
(indeed, the BTZ solution can be obtained as the quotient of AdS$_3$ by a discrete
group of invariances  \cite{Banados:1992gq}). The identification between the 
coordinate $r$ in Eq.~\eqref{eq:BTZ_metric} and $\mu$ is given 
by\footnote{Since the calculations in this section are mostly numerical, 
we use different unit conventions, in which 
$r, r_h, t, \tau, \delta$ are dimensionless. All dimensional factors are carried out by $L$.}
\beq
\tan \mu = \pm \cosh (r_h t)  \sqrt{\le \frac{r}{r_h} \ri^2 -1} \, .
\label{eq:orbifold_condition}
\eeq
This is responsible for the time-dependence of the final configuration,
because  $\mu$ is function both of $t$ and $r$. We can introduce the Kruskal coordinates
\beq
\label{eq:Kruskal_BTZ}
V=e^{r_h(t+r_*)} \, , \qquad
U=-e^{-r_h(t-r_*)} \, ,  \qquad
\eeq
with tortoise coordinate
\beq
\label{eq:tortoise}
r_* (r)= \frac{1}{2r_h} \log \le \frac{r-r_h}{r+r_h} \ri \, .
\eeq
which can be compactified by introducing
\beq
\label{eq:Kruskal_BTZ_2}
V= \tan w_1 \, , \qquad
U= \tan w_2 \, ,
\eeq
and finally we go back from null coordinates to timelike and spacelike ones defining
\beq
\label{eq:second_change_BTZ}
\tau = w_1 + w_2 \, , \qquad
\mu= w_1 - w_2 \, .
\eeq
Using the relation \eqref{eq:orbifold_condition}, we obtain \cite{Bak:2007jm, Bak:2007qw}
\beq
ds^2_3 = L^2 f(\mu) \le -d\tau^2 + d\mu^2 +r_h^2 \cos^2 \tau d \theta^2 \ri \, .
\label{eq:metric1_Janus_BTZ}
\eeq
Changing variable as in Eq.~\eqref{change-vars-y-mu}, we get
\beq
ds^2_3 = L^2 dy^2 + L^2 f(y) \le - d\tau^2 + r_h^2 \cos^2 \tau d\theta^2 \ri \, ,
\label{eq:metric2_Janus_BTZ}
\eeq
with the same dilaton solution given in Eq.~\eqref{eq:function_f_and_dilaton}.
The Penrose diagram is shown in Fig.~\ref{fig:Penrose_Janus_BTZ}.
Note the appearance of a "shadow region",
which is causally disconnected from both the boundaries.
For $\gamma=0$ we recover the BTZ solution,
which has a square Penrose diagram with no shadow region.

\subsection{Volume at vanishing boundary time}
\label{sect-volume_vanishing_boundary_time_JBTZ_time_dep}

The time-dependent Janus deformation of the BTZ black hole has two disconnected boundaries.
In order to analyze the structure of the UV divergences of the volume, it is sufficient to consider the case where both the boundary times vanish $\tau_L = \tau_R =0,$ which implies that the extremal surface sits on a time slice at constant $\tau=0.$ 
This is interpreted as the complexity of formation of the Janus deformation starting from the pure BTZ background.
The extremal volume reads
\beq
\mathcal{V}_0 (\gamma) = 2 L^2 \, r_h  \int_0^{2 \pi} d \theta \int_0^{y^*} dy \, \sqrt{f(y)} \, ,
\label{eq:intermediate_volume_JBTZ_nonstatic}
\eeq
where introduce a UV cutoff at $y=y^*,$ which we explictly determine below.

In order to compute the value of the UV cutoff consistently with the FG expansion,
 we notice from \cite{Ugajin:2014nca} that near the conformal boundary 
 the metric approaches AdS space in Poincaré coordinates with radial direction $\xi$ and time coordinate $t$ if we make the identification
\beq
\xi= \frac{2}{r_h \, (1-2\gamma^2)^{\frac{1}{4}}} e^{-|y|} \, \cosh \le r_h t \ri \, .
\label{eq:FG_time_dep_JBTZ}
\eeq
The compact time $\tau$ and the non-compact one $t$ are related at the boundary by the relation
\beq
\tanh (r_h t) = \sin \tau \, .
\label{eq:change_coordinates_t_tau}
\eeq
If we set $\xi= \delta$ and $t=0,$  we get
\beq
y^* = - \log \le \frac{r_h (1-2\gamma^2)^{\frac{1}{4}}}{2} \, \delta \ri  \, .
\label{eq:ystar_time_dep_JBTZ_vanishing_time}
\eeq
The result for the volume in Eq.~\eqref{eq:intermediate_volume_JBTZ_nonstatic}  is
\beq
\mathcal{V}_0 (\gamma)   
 = 2 \pi L^2 \le \frac{2}{\delta} + r_h \eta(\gamma)
     \ri \, ,
\label{eq:volume_time_vanishing_time_dep_JBTZ}
\eeq
where  $\eta(\gamma)$ is defined in Eq.~\eqref{eta-gamma-eq}.
 The difference between this volume and the volume of 
the standard BTZ solution (obtained for $\gamma=0$) is
\beq
\Delta \mathcal{V}_0 (\gamma) \equiv \mathcal{V}_0 (\gamma) - \mathcal{V}_0 (0) = 
2 \pi  L^2 r_h \eta(\gamma) \, .
\label{Delta_V_time_zero-time-dependent-BTZ}
\eeq
This quantity is indeed positive (see figure \ref{eta-gamma}).

\subsection{Growth rate of the volume}
\label{sect-growth_rate_volume_analytic_part}

In this section we will address the time evolution of the ERB. 
In principle, the extremal surface can be specified by a function $\tau(y,\theta)$.
Due to the cylindrical symmetry, the extremal $\tau$  does not depend on $\theta$, i.e.,
\beq
\tau(y, \theta) = \tau (y) \, .
\eeq
The volume functional can be evaluated from the metric 
in Eq.~(\ref{eq:metric2_Janus_BTZ}), giving
\beq
\mathcal{V} (\gamma) = 4 \pi L^2 r_h \int_{0}^{y_{\rm max}} dy \, 
f (y) \cos \left[ \tau(y) \right] \, \sqrt{1-f(y) \dot{\tau}^2} \, ,
\label{eq:extremal_volume_tau_coord_JBTZ}
\eeq
 where the dot denotes derivative with respect to $y$, and 
 $y_{\rm max}$ will be fixed below in terms of an appropriate UV cutoff.
The Euler-Lagrange equation resulting from  Eq.~(\ref{eq:extremal_volume_tau_coord_JBTZ}) is
\beq
 2 \sin \tau - 3 \cos \tau \, f' (y) \dot{\tau} + 
2 \cos \tau \, f(y) f'(y) \dot{\tau}^3  - 2 f(y) \le \sin \tau \, \dot{\tau}^2 + \cos \tau \, \ddot{\tau} \ri  = 0 \, .
\label{eq:gauge_fixed_diff_eq_Euler_Lagrange}
\eeq
We then need to specify  appropriate boundary conditions.
We consider for simplicity the case where the boundary times satisfy 
\beq
\tau_B \equiv \tau_L = \tau_R  \, .
\eeq
In this case, the geometry is still symmetric between the left and right sides of the Penrose diagram: continuity implies that there exists a turning point for the extremal slice.
For these reasons, we can impose the conditions
\beq
\tau (y_{\rm max}) =\tau (-y_{\rm max}) =\tau_B \, , \qquad
 \frac{d \tau}{dy} \Big|_{y=0} = 0 \, . 
 \label{eq:boundary_conditions_Eulero_Lagrange}
\eeq
In particular, the turning point is identified by the condition
\beq
\tau_{\rm min} \equiv \tau (y=0) \, .
\eeq
Changing the boundary time $\tau_B$, the value of $\tau_{\rm min}$ gets modified.

\subsection{Numerical solutions and time dependence of the volume}
\label{sect-numerical_solutions_JBTZ_time_dep}

We determine the numerical solutions to the Euler-Lagrange equations given in \eqref{eq:gauge_fixed_diff_eq_Euler_Lagrange}.
This means that we find the extremal slices representing the evolution of the ERB, and we also compute numerically the integral which gives their volume.

\subsubsection*{Extremal slices}

\begin{figure}[ht]
    \centering
    \def\svgwidth{\columnwidth}
    \scalebox{1}{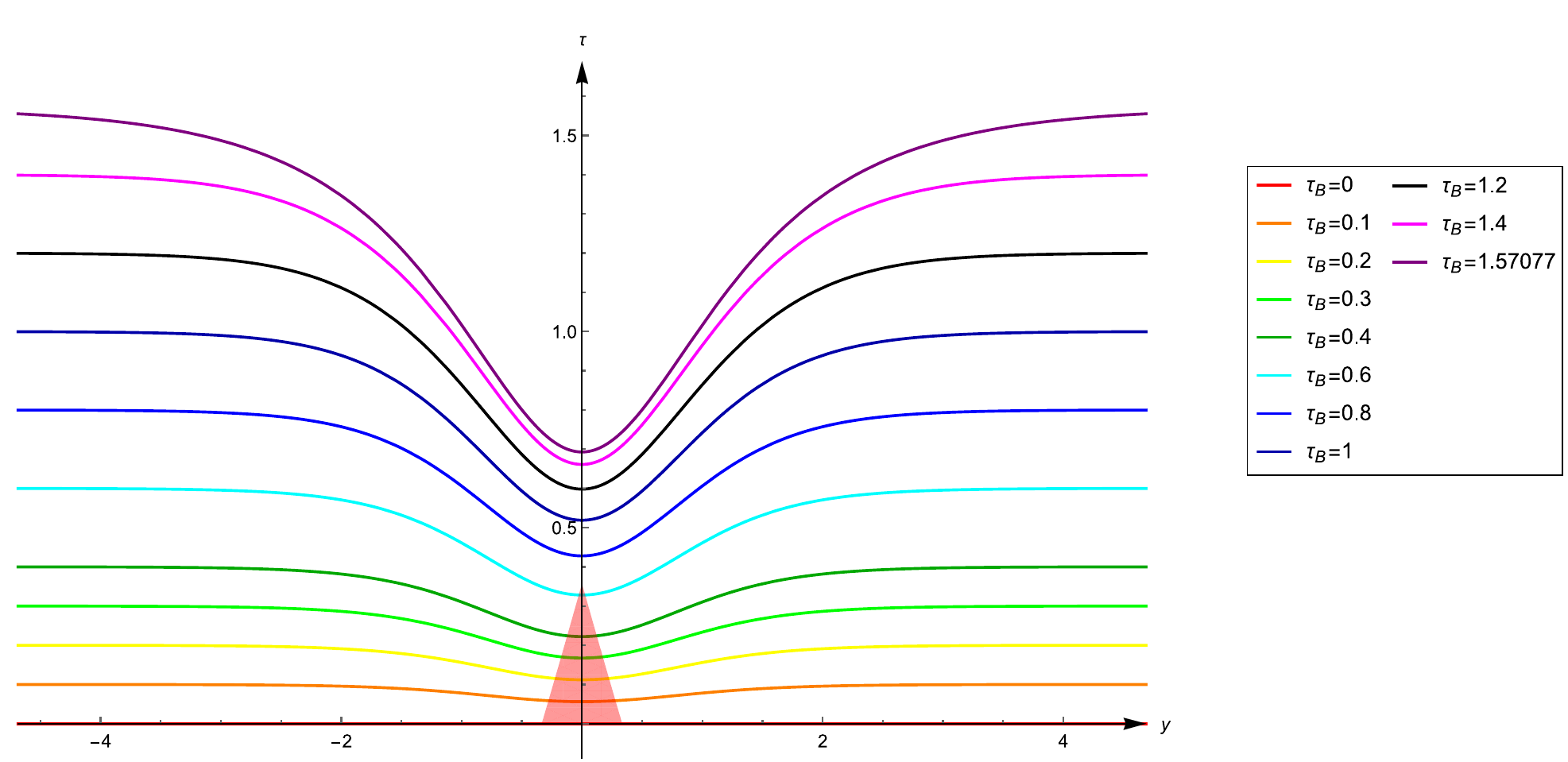}
    \caption{Plot of the numerical solutions $\tau=\tau(y)$ for different values of the boundary time $\tau_B. $ The configuration is taken to be symmetric, i.e., $\tau_L=\tau_R=\tau_B$
   and the solutions have a turning point at $y=0,$ which corresponds to $\tau_{\rm min}.$
All the extremal slices are computed by putting $\gamma=0.5, r_h=2.0$ and $\delta=10^{-4}.$ 
The solution with $\tau_B=0$ corresponds to the case studied in Section \ref{sect-volume_vanishing_boundary_time_JBTZ_time_dep}. 
The shadow region is shown as the red triangle in the picture.}
\label{fig-numerical_solutions_FGcutoff_time_dep_JBTZ}
\end{figure}

An appropriate choice of the UV cutoff is naturally determined by Eq.~\eqref{eq:FG_time_dep_JBTZ} evaluated at $\xi=\delta.$
We find
\beq
y_{\rm max} (\tau_B) =  \log \le  \frac{2}{r_h \,\delta} \, \frac{1}{\cos \tau_B (1-2\gamma^2)^{\frac{1}{4}}}  \ri  \, .
\eeq
This choice corresponds to the FG coordinates such that the metric on the boundary is the BTZ black hole solution, and indeed the cutoff is time-dependent as a result of the non-stationarity of the Janus deformation.
Imposing the boundary conditions \eqref{eq:boundary_conditions_Eulero_Lagrange} at these values of the $y$ coordinate, we get the numerical solutions  in Fig.~\ref{fig-numerical_solutions_FGcutoff_time_dep_JBTZ}.

One may wonder which extremal volumes have a turning point that sits inside the \emph{shadow region}.
This part of the spacetime is determined by the intersection of the curves at constant value of the null coordinates
\beq
w_1 = \frac{1}{2} \le \tau + \mu \ri \, , \qquad
w_2 = \frac{1}{2} \le \tau - \mu \ri \, .
\eeq
If we impose the condition that these null lines pass through the points 
\beq
(\mu, \tau) = \le -\mu_0, - \frac{\pi}{2} \ri \, , \qquad
(\mu, \tau) = \le \mu_0,- \frac{\pi}{2} \ri \, ,
\eeq
then the curves that determine the shadow region for positive $\tau$ are given by
\beq
-\mu + \mu_0 = \tau + \frac{\pi}{2} \, , \qquad
\mu + \mu_0 = \tau + \frac{\pi}{2} \, .
\eeq
The corresponding equation in the $(y,\tau)$ coordinate system is obtained after performing the transformation
\beq
\tanh y = \mathrm{sn} \, \le \alpha_+\mu \Big| m \ri 
 \, , \qquad
 m = \frac{\alpha_-^2}{\alpha_+^2} \, , \qquad
\alpha_{\pm}^2 = \frac{1\pm\sqrt{1-2\gamma^2}}{2} \, , 
\eeq
and using the definition
\beq
\mu_0 = \frac{1}{\alpha_+} \mathbb{K}\le \frac{\alpha_-^2}{\alpha_+^2} \ri  \geq \frac{\pi}{2} \, .
\eeq
The shadow region is coloured in red in Fig.~\ref{fig-numerical_solutions_FGcutoff_time_dep_JBTZ}.

\subsubsection*{UV divergencies of the volume}

We numerically evaluate the integral corresponding to the extremal
 volume \eqref{eq:extremal_volume_tau_coord_JBTZ}, with  $\tau(y)$ 
  numerically determined by solving the differential equation \eqref{eq:gauge_fixed_diff_eq_Euler_Lagrange}
   with boundary conditions \eqref{eq:boundary_conditions_Eulero_Lagrange}.
We find that the volume is a monotonically increasing function of time.
A typical feature of black hole solutions like the BTZ background is that the divergences are time-independent, while the finite part brings the information about the time evolution of the system.
This can be heuristically understood from the fact that the time dependence is encoded in the turning point $\tau_{\rm min}$, and since it is determined at $y=0,$ which is far away from the boundary, it carries no UV modes.
We suspect that the same phenomenon occurs for this non-stationary black hole deformation, since the choice of the UV cutoff determined from the FG expansion is such that we recover the static BTZ background at the boundary.

In order to investigate  the dependence of the volume from the UV cutoff $\delta$, we consider
a fixed boundary time $\tau_B$ and we choose different values of $\delta$ scaling as
\beq
\delta_i = 10^{-\frac{i+2}{2}} \, ,
\label{eq:succession_UV_cutoff_JBTZ}
\eeq
with $i \in \lbrace 1, \dots , 10 \rbrace .$
If we assume that the leading divergence of the extremal volume scales as
\beq
 \mathcal{V}_i \equiv \mathcal{V} (\delta_i, \tau_B)  = \frac{4 \pi}{\delta_i} + \mathcal{O} (\delta_i^0) \, ,
\eeq
see Eq.~(\ref{eq:volume_time_vanishing_time_dep_JBTZ}),
this would imply that the ratio between the volume evaluated for two consecutive values of 
$\delta_i$
 is constant
\beq
\frac{\mathcal{V}_{i+1}}{\mathcal{V}_i} = \sqrt{10} \simeq 3.162 \, .
\eeq
We tested this behaviour numerically
  for various choices of the boundary time.
We find a strong numerical evidence that the leading divergence indeed scales as $\delta^{-1},$
 see Table~\ref{tab:time_dep_BTZ}.

\begin{table}[ht]     
\begin{center}    
\begin{tabular}  {|c|c|c|} \hline Boundary time $\tau_B$ & Ratio $\mathcal{V}_{i+1}/ \mathcal{V}_i$  \\ \hline
0.2 & $3.159 \pm 0.006 $  \\ 
0.4 & $3.158 \pm 0.007 $  \\ 
0.6 & $3.157 \pm 0.010 $ \\ 
0.8 & $3.154 \pm 0.014 $ \\ 
1.0 & $3.151 \pm 0.020 $ \\ 
1.2 & $3.144 \pm 0.029 $ \\
\hline
\end{tabular}   
\caption{Numerical value of the ratios $\mathcal{V}_{i+1}/ \mathcal{V}_i$ evaluated at various boundary times $\tau_B.$
We determined the mean value and the standard deviation from the results obtained by varying $i \in \lbrace 1, \dots  10 \rbrace$ in the function $\delta = 10^{-\frac{i+2}{2}}.$
The results must be compared with the analytic expectation $\sqrt{10} \simeq 3.162.$
\label{tab:time_dep_BTZ}}
\end{center}
\end{table}

We also notice that these ratios have a bigger error and deviate more from the analytic result as we increase the boundary time.
This phenomenon can be  explained by the fact that the finite term increases with time, and therefore produces a small deviation of the total volume from the contribution coming only from the leading divergence.
One can verify that at $\tau_B=0$ the analytic result in  Eq.~(\ref{eq:volume_time_vanishing_time_dep_JBTZ}) for the BTZ background is recovered with great accuracy.

\subsubsection*{Time dependence of the volume}

\begin{figure}[ht]
    \centering
   \def\svgwidth{\columnwidth}
    \scalebox{0.7}{
\begingroup%
  \makeatletter%
  \providecommand\color[2][]{%
    \errmessage{(Inkscape) Color is used for the text in Inkscape, but the package 'color.sty' is not loaded}%
    \renewcommand\color[2][]{}%
  }%
  \providecommand\transparent[1]{%
    \errmessage{(Inkscape) Transparency is used (non-zero) for the text in Inkscape, but the package 'transparent.sty' is not loaded}%
    \renewcommand\transparent[1]{}%
  }%
  \providecommand\rotatebox[2]{#2}%
  \newcommand*\fsize{\dimexpr\f@size pt\relax}%
  \newcommand*\lineheight[1]{\fontsize{\fsize}{#1\fsize}\selectfont}%
  \ifx\svgwidth\undefined%
    \setlength{\unitlength}{518bp}%
    \ifx\svgscale\undefined%
      \relax%
    \else%
      \setlength{\unitlength}{\unitlength * \real{\svgscale}}%
    \fi%
  \else%
    \setlength{\unitlength}{\svgwidth}%
  \fi%
  \global\let\svgwidth\undefined%
  \global\let\svgscale\undefined%
  \makeatother%
  \begin{picture}(1,0.64671815)%
    \lineheight{1}%
    \setlength\tabcolsep{0pt}%
    \put(0,0){\includegraphics[width=\unitlength,page=1]{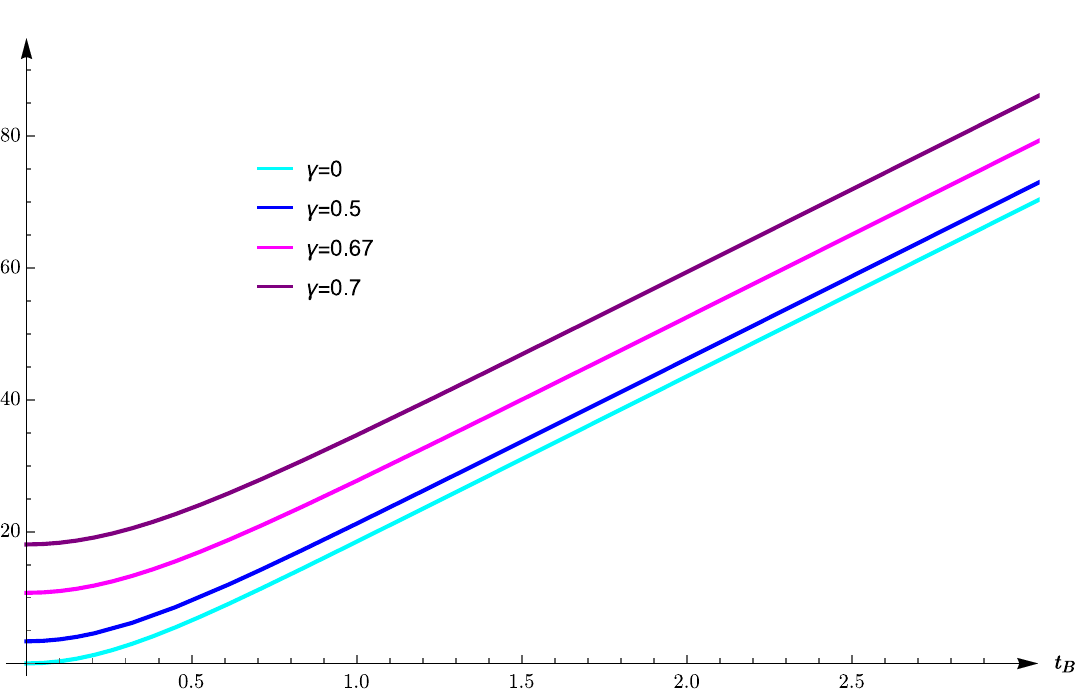}}%
    \put(0.01480439,0.62723869){\makebox(0,0)[lt]{\lineheight{1.25}\smash{\begin{tabular}[t]{l}$\mathcal{V}_{ren}$\end{tabular}}}}%
    \put(0,0){\includegraphics[width=\unitlength,page=2]{volume_growth.pdf}}%
    \put(0.964623,0.01947946){\makebox(0,0)[lt]{\lineheight{1.25}\smash{\begin{tabular}[t]{l}$t_B$\end{tabular}}}}%
    \put(0,0){\includegraphics[width=\unitlength,page=3]{volume_growth.pdf}}%
    \put(0.27894592,0.48498628){\makebox(0,0)[lt]{\lineheight{1.25}\smash{\begin{tabular}[t]{l}$\gamma=0$\end{tabular}}}}%
    \put(0.27886271,0.4457184){\makebox(0,0)[lt]{\lineheight{1.25}\smash{\begin{tabular}[t]{l}$\gamma=0.5$\end{tabular}}}}%
    \put(0.27886271,0.41143454){\makebox(0,0)[lt]{\lineheight{1.25}\smash{\begin{tabular}[t]{l}$\gamma=0.67$\end{tabular}}}}%
    \put(0.27808354,0.37225465){\makebox(0,0)[lt]{\lineheight{1.25}\smash{\begin{tabular}[t]{l}$\gamma=0.7$\end{tabular}}}}%
    \put(0,0){\includegraphics[width=\unitlength,page=4]{volume_growth.pdf}}%
  \end{picture}%
\endgroup%
}
    \caption{Time dependence of the volume of the solutions depicted in Fig.~\ref{fig-numerical_solutions_FGcutoff_time_dep_JBTZ} for different values of $\gamma$ , with $\delta=10^{-4}$. The volumes $\mathcal{V}_{\rm ren}$ are renormalized after subtracting the divergence $4 \pi/ \delta .$}
\label{fig-numerical_volume_BTZ_gamma}
\end{figure}

\begin{figure}[ht]
    \centering
    \def\svgwidth{\columnwidth}
    \scalebox{0.75}{
\begingroup%
  \makeatletter%
  \providecommand\color[2][]{%
    \errmessage{(Inkscape) Color is used for the text in Inkscape, but the package 'color.sty' is not loaded}%
    \renewcommand\color[2][]{}%
  }%
  \providecommand\transparent[1]{%
    \errmessage{(Inkscape) Transparency is used (non-zero) for the text in Inkscape, but the package 'transparent.sty' is not loaded}%
    \renewcommand\transparent[1]{}%
  }%
  \providecommand\rotatebox[2]{#2}%
  \newcommand*\fsize{\dimexpr\f@size pt\relax}%
  \newcommand*\lineheight[1]{\fontsize{\fsize}{#1\fsize}\selectfont}%
  \ifx\svgwidth\undefined%
    \setlength{\unitlength}{518bp}%
    \ifx\svgscale\undefined%
      \relax%
    \else%
      \setlength{\unitlength}{\unitlength * \real{\svgscale}}%
    \fi%
  \else%
    \setlength{\unitlength}{\svgwidth}%
  \fi%
  \global\let\svgwidth\undefined%
  \global\let\svgscale\undefined%
  \makeatother%
  \begin{picture}(1,0.65637066)%
    \lineheight{1}%
    \setlength\tabcolsep{0pt}%
    \put(0,0){\includegraphics[width=\unitlength,page=1]{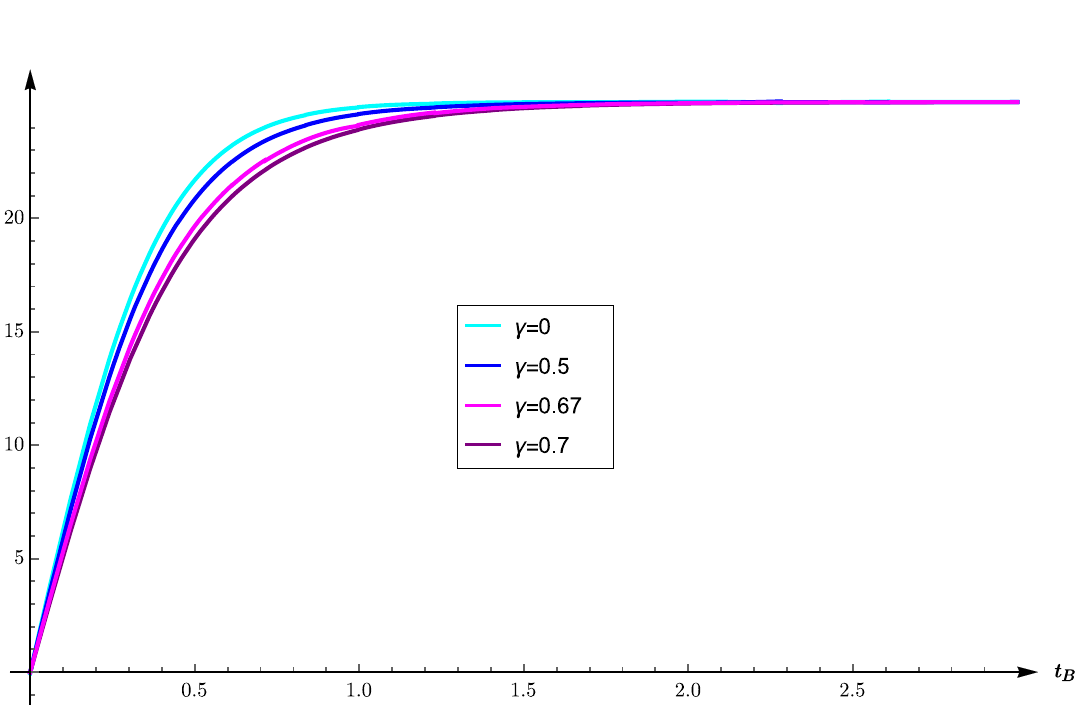}}%
    \put(0.00080244,0.6157836){\makebox(0,0)[lt]{\lineheight{1.25}\smash{\begin{tabular}[t]{l}$\frac{d\mathcal{V}_{ren}}{dt_B}$\end{tabular}}}}%
    \put(0,0){\includegraphics[width=\unitlength,page=2]{volume_rate.pdf}}%
    \put(0.47394987,0.34531561){\makebox(0,0)[lt]{\lineheight{1.25}\smash{\begin{tabular}[t]{l}$\gamma=0$\end{tabular}}}}%
    \put(0.4729014,0.31171134){\makebox(0,0)[lt]{\lineheight{1.25}\smash{\begin{tabular}[t]{l}$\gamma=0.5$\end{tabular}}}}%
    \put(0.47185476,0.27559001){\makebox(0,0)[lt]{\lineheight{1.25}\smash{\begin{tabular}[t]{l}$\gamma=0.67$\end{tabular}}}}%
    \put(0.47184313,0.23787544){\makebox(0,0)[lt]{\lineheight{1.25}\smash{\begin{tabular}[t]{l}$\gamma=0.7$\end{tabular}}}}%
    \put(0,0){\includegraphics[width=\unitlength,page=3]{volume_rate.pdf}}%
    \put(0.97030984,0.02372422){\makebox(0,0)[lt]{\lineheight{1.25}\smash{\begin{tabular}[t]{l}$t_B$\end{tabular}}}}%
  \end{picture}%
\endgroup%
}
  	\caption{Time derivative of the extemal volume of the time dependent Janus BTZ deformation for various values of $\gamma$.}
  	\label{fig_volume_BTZ_subtraction_gamma}
\end{figure}

In order to show the time dependence of the solution from the physical time $t_B$, we use the change of coordinates, valid close to the boundary, given by Eq.~\eqref{eq:change_coordinates_t_tau}.
After subtracting the divergence $4\pi/\delta,$ which is the same for all values of $\gamma$,
we show in Fig.~\ref{fig-numerical_volume_BTZ_gamma} the time dependence of the volume $\mathcal{V}_{\rm ren}$ for both the Janus deformation, plotted for different values of $\gamma$, and the pure BTZ background $\gamma=0$.
The result is a monotonically increasing function of time, which reaches a linear growth for late times. We also depict the growth-rate of the volume in Fig.~\ref{fig_volume_BTZ_subtraction_gamma}. 
We numerically find that the rates of growth of the volume for late times are the same (within numerical uncertainties) for all the solutions with various $\gamma.$

\begin{figure}[ht]
    \centering
    \def\svgwidth{\columnwidth}
    \scalebox{0.75}{
\begingroup%
  \makeatletter%
  \providecommand\color[2][]{%
    \errmessage{(Inkscape) Color is used for the text in Inkscape, but the package 'color.sty' is not loaded}%
    \renewcommand\color[2][]{}%
  }%
  \providecommand\transparent[1]{%
    \errmessage{(Inkscape) Transparency is used (non-zero) for the text in Inkscape, but the package 'transparent.sty' is not loaded}%
    \renewcommand\transparent[1]{}%
  }%
  \providecommand\rotatebox[2]{#2}%
  \newcommand*\fsize{\dimexpr\f@size pt\relax}%
  \newcommand*\lineheight[1]{\fontsize{\fsize}{#1\fsize}\selectfont}%
  \ifx\svgwidth\undefined%
    \setlength{\unitlength}{518bp}%
    \ifx\svgscale\undefined%
      \relax%
    \else%
      \setlength{\unitlength}{\unitlength * \real{\svgscale}}%
    \fi%
  \else%
    \setlength{\unitlength}{\svgwidth}%
  \fi%
  \global\let\svgwidth\undefined%
  \global\let\svgscale\undefined%
  \makeatother%
  \begin{picture}(1,0.64671815)%
    \lineheight{1}%
    \setlength\tabcolsep{0pt}%
    \put(0,0){\includegraphics[width=\unitlength,page=1]{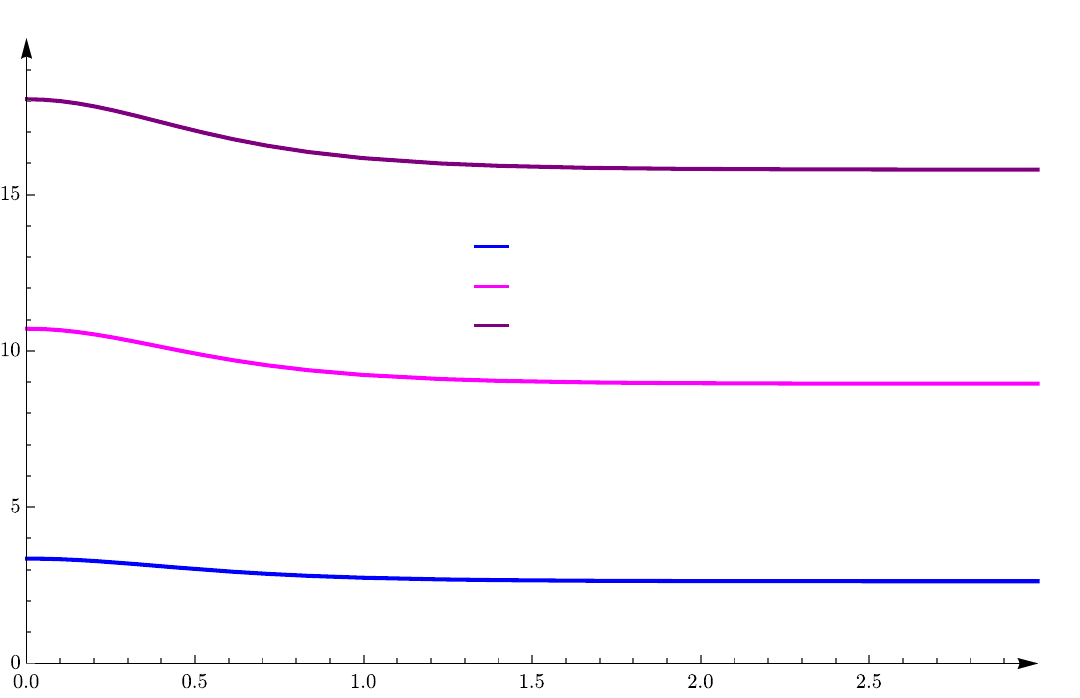}}%
    \put(0.00779179,0.62645951){\makebox(0,0)[lt]{\lineheight{1.25}\smash{\begin{tabular}[t]{l}$\Delta\mathcal{V}$\end{tabular}}}}%
    \put(0.96696054,0.02025866){\makebox(0,0)[lt]{\lineheight{1.25}\smash{\begin{tabular}[t]{l}$t_B$\end{tabular}}}}%
    \put(0.47841559,0.40806856){\makebox(0,0)[lt]{\lineheight{1.25}\smash{\begin{tabular}[t]{l}$\gamma=0.5$\end{tabular}}}}%
    \put(0.4801059,0.37035903){\makebox(0,0)[lt]{\lineheight{1.25}\smash{\begin{tabular}[t]{l}$\gamma=0.67$\end{tabular}}}}%
    \put(0.48010593,0.33451684){\makebox(0,0)[lt]{\lineheight{1.25}\smash{\begin{tabular}[t]{l}$\gamma=0.7$\end{tabular}}}}%
    \put(0,0){\includegraphics[width=\unitlength,page=2]{volume_sub.pdf}}%
  \end{picture}%
\endgroup%
}
  	\caption{Plot of the quantity $\Delta  \mathcal{V}(t_B) $ defined in eq. (\ref{volume-meno-BTZ})
	for various values of $\gamma$. When $\gamma=0,$ this function is vanishing by construction.}
  	\label{fig_volume_BTZ_subtraction_gamma-2}
\end{figure}

According to \cite{Bak:2007qw}, the time-dependent Janus deformation of the BTZ solution
describes the thermalisation of an out-of equilibrium system, with a late-time temperature
given by
\beq
T = \frac{r_h}{2 \pi \, L} \, .
\eeq
In addition, the area of the horizon is given by \cite{Bak:2007jm}
\beq
\mathcal{A} (\tau) = 2 \pi L \,  r_h \frac{\alpha_+ \sin \le \pi/2 - \tau \ri}{\mathrm{sn}\, \le \alpha_+ \le \pi/2 - \tau \ri, k \ri} 
\underset{t_B \rightarrow \infty}{\longrightarrow} 2 \pi L \, r_h \, .
\eeq
Hence, at late times
\beq
 TS = \frac{r_h^2}{4 G} \, ,
\eeq
which is the same value that one gets for the BTZ background.
We numerically checked,  for the values of $\gamma$ shown in fig. \ref{fig_volume_BTZ_subtraction_gamma-2},
that the asymptotic rate of the volume is
\beq
\frac{1}{4 \pi G \, L^2} \, \lim_{t_B \rightarrow \infty} \frac{d \mathcal{V}_{\rm ren}}{d t_B} = T S \, ,
\eeq
which matches with the volume rate at late times found in \cite{Carmi:2017jqz} for the BTZ case.

We remark that the divergence of the volume is universal because it does not depend on $\gamma$.
For this reason, it is meaningfull to compare the finite part of the volume with different values of the Janus deformation.  
We define
\beq
\Delta  \mathcal{V}(\gamma, t_B) =  \mathcal{V}(\gamma, t_B)- \mathcal{V}(0, t_B) \, ,
\label{volume-meno-BTZ}
\eeq
which is plotted (as a function of $t_B$ for different $\gamma$)
 in figure \ref{fig_volume_BTZ_subtraction_gamma-2}.
At $t_B=0$, we have that $\Delta  \mathcal{V}=\Delta  \mathcal{V}_0$, 
given by eq. (\ref{Delta_V_time_zero-time-dependent-BTZ}).
The quantity $\Delta  \mathcal{V}_0$  is positive and, in particular, 
for $\gamma \to \sqrt{2}/{2}$ it diverges. Since the volume rate is 
a decreasing function of $\gamma$,  we have that $\Delta  \mathcal{V}(\gamma, t_B) $
is a decreasing function of time. The asymptotic rate does not depend
on $\gamma$ and so $\Delta  \mathcal{V}$ approaches  a constant at late times.

For $\gamma \neq 0$  the system at $t_B=0$ starts in a out-of-equilibrium state
and then thermalizes at late times. The resulting complexity rate is lower
compared to the $\gamma=0$ case: the computational power
gets decreased by the time-dependent perturbation which brings the 
system out of equilibrium.
The initial $\Delta  \mathcal{V}_0$ is partially washed out
at later times, but it does not approach  zero asymptotically.
It is surprising that the late-time volume rate is universal,
in spite of the fact that different boundary values of the dilaton
are dual to theories with different couplings.

\subsection{Contribution of the shadow region}

The consistency of  holographic entanglement entropy  with causality  \cite{Headrick:2014cta}
imposes the requirement that the HRT surface of a given boundary region $\mathcal{A}$ lies
in the causal shadow of $\mathcal{A}$. When the region $\mathcal{A}$
coincides with the left or right boundary of the Kruskal diagram, the 
causal shadow coincides with the shadow region in figure \ref{fig:Penrose_Janus_BTZ}.
The HRT surface of one of the boundaries of the time-dependent Janus BH 
lies at the center of the Penrose diagram and so it satisfies the causality bound.

\begin{figure}[ht]
\centering
\includegraphics[scale=0.45]{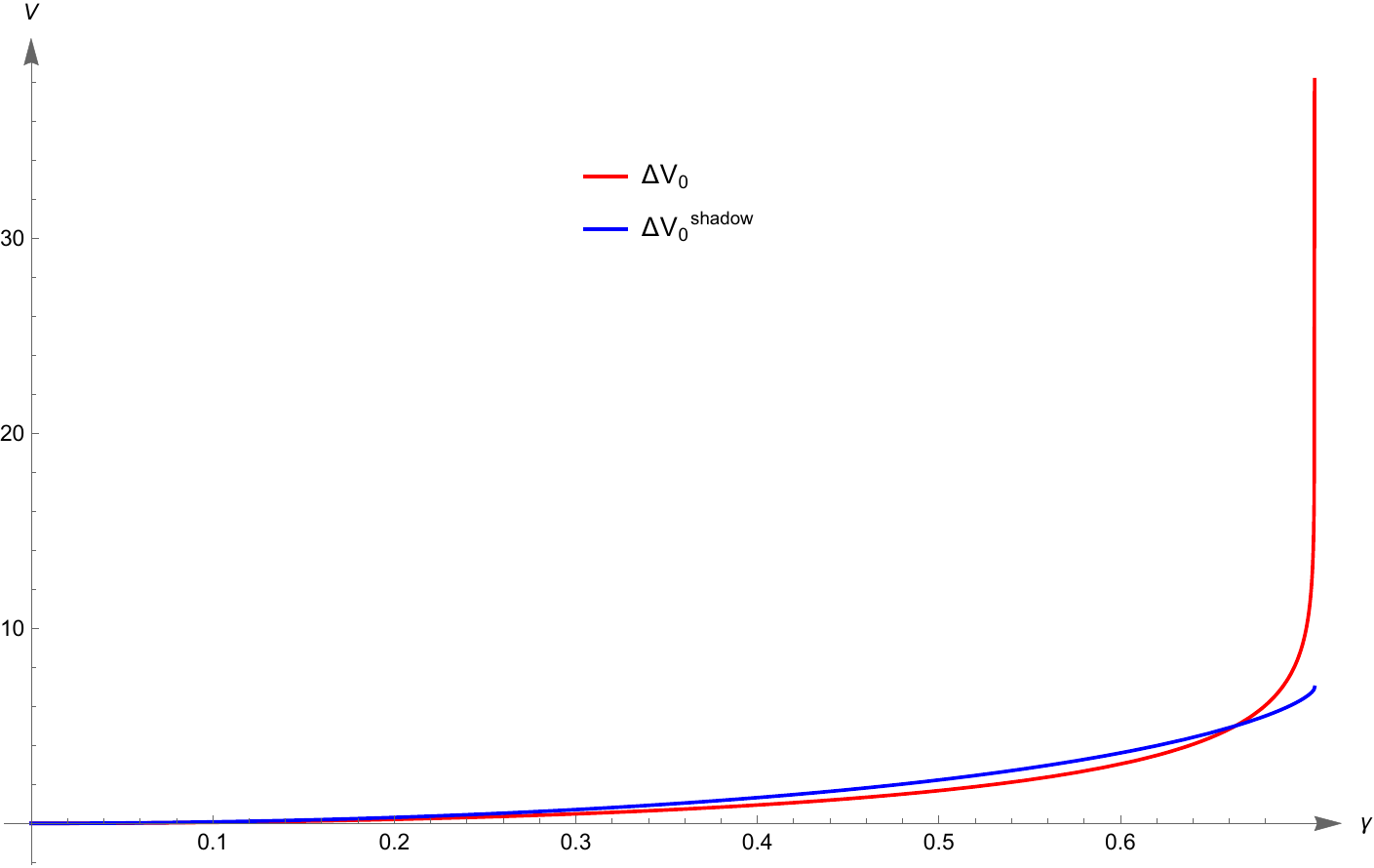}
\caption{Volumes $\Delta \mathcal{V}_0$ and $\Delta \mathcal{V}_0^{\rm shadow} $ 
as a fuction of $\gamma$. The numerical value $r_h=1$ is chosen.}
\label{fig-contribution_shadow_region}
\end{figure}

Given the importance of the causal shadow for holographic entanglement entropy,
it is natural to wonder about the role of the shadow region in the Complexity=Volume conjecture.
The extremal volume  at $t=0$ actually crosses horizontally the entire shadow causal diamond 
in figure \ref{fig:Penrose_Janus_BTZ}. It is then interesting to compute 
the part of the volume which is included inside the shadow region at $t=0$.
This can be achieved by  restricting  to the interval $\mu \in [0, \mu_0 - \frac{\pi}{2}]$
the integral in eq. (\ref{eq:intermediate_volume_JBTZ_nonstatic}). A direct evaluation gives
\beq
\Delta \mathcal{V}_0^{\rm shadow} (\gamma) = 2 L^2 r_h \int_0^{2 \pi} d \theta \, 
\int_0^{\mathrm{arctanh} \left[ \mathrm{sn} \le \alpha_+ \le \mu_0 - \frac{\pi}{2} \ri |m \ri \right]} dy \, \sqrt{f(y)} \, .
\eeq
It is interesting to compare $\Delta \mathcal{V}_0^{\rm shadow} (\gamma)$  with $\Delta \mathcal{V}_0(\gamma)$
in eq. (\ref{Delta_V_time_zero-time-dependent-BTZ}), see fig.~\ref{fig-contribution_shadow_region}.
For small $\gamma$, the dominant contribution to  $\Delta \mathcal{V}_0(\gamma)$ comes indeed from the shadow region.
This is not true for $\gamma$ approaching to the linear dilaton regime ($\gamma \rightarrow 1/\sqrt{2}$):
in this limit, the quantity $\Delta \mathcal{V}_0(\gamma)$ diverges, 
while the contribution arising from the shadow region is always finite, reaching the value $\pi^2 r_h / \sqrt{2}$. 
The linear dilaton limit lies in the strongly coupled regime from the bulk perspective,
and we expect that in this limit our classical holographic calculations are not trustable.

At larger $t_B$, the extremal volume surface tends to escape from the shadow region.
From fig.~\ref{fig-numerical_solutions_FGcutoff_time_dep_JBTZ}, we can check that for $\gamma=0.5$ 
and for $\tau_B>  \tau_{B0} \approx 0.6,$ the extremal volume starts to sit completely outside the shadow region.
The escape of the extremal volume brane from the shadow region
does not seem to affect the time dependence of complexity 
in a particularly dramatic way. As can be checked
from figures \ref{fig-numerical_volume_BTZ_gamma} and \ref{fig_volume_BTZ_subtraction_gamma},
both the renormalised complexity and the  complexity rate 
remain continuous and no special feature
seems to appear for  $\tau_B= \tau_{B0}$.


\section{Conclusions}
\label{sect:conclu}

In this paper we studied subregion volume complexity 
for an interval of length $l$ centered around the AdS$_3$ 
Janus interface. The contribution to the complexity due to the defect
is log divergent in the cutoff $\delta$, i.e.
\beq
\Delta \mathcal{C}(l,\gamma)  =
\frac{2}{3} c \,  \eta(\gamma) \,
\log \le    \frac{l}{\,  \delta} \ri
 + \text{finite terms}  \, ,
 \label{result-janus-Tzero-2}
\eeq
where $ \eta(\gamma) $ is defined in terms of elliptic integrals, see eq. (\ref{eta-gamma-eq})
and  figure \ref{eta-gamma}. 
We performed the calculation with three different regularizations:
Fefferman-Graham, single and double cutoff. 
The three methods give the same result for the log divergent term in eq. (\ref{result-janus-Tzero-2}),
while the finite terms depend on the choice of regulator.
We compared the result with the AdS$_3$ Randall-Sundrum and the BCFT models,
see section \ref{comparison-other-geometries} for details. 
In all these models the leading divergence in $\Delta \mathcal{C}$ is logarithmic.
This divergence  is not related in an universal way to the boundary 
entropy $\Delta S$, even if the two quantities share a similar qualitative behaviour.

In absence of the interface, the subregion volume complexity for the BTZ background 
 is topologically protected by the Gauss-Bonnet theorem  \cite{Abt:2017pmf} and, therefore, does not depend on temperature. 
 This is no longer true in the presence of the Janus interface.
 We computed the difference $\Delta \mathcal{C}_T$ between 
 the finite and the zero temperature subregion complexity 
 using both single and double cutoff regularizations.
This is a finite quantity, whose temperature dependence does not depend on
the regularization, i.e.
 \beq
\Delta \mathcal{C}_T (T,l,\gamma)  =
\frac23 \, c   \, \eta(\gamma)    \,  \log \left[ \frac{2}{\pi   l T} \tanh \left(\frac{\pi l T}{2 } \right) \right]  
+ \dots \, ,
\eeq
where the $\dots$ correspond to finite terms which are function just of $\gamma$,
which parameterizes the excursion
of the dilaton between the two sides of the interface.
This is a decreasing function of temperature, see figure \ref{comple-temperatura}.
It would be interesting to compute this quantity also in the RS and in
the BCFT models, in order to see if there is some universality property.

As the $l \to \infty$ limit covers the whole space, we can define the total complexity
by the $l \to \infty$ limit of the subregion complexity.
The complexity of formation for the defect is then defined
as the difference between the total complexity 
and its value at $\gamma=0$.
At zero temperature, the result is given by Eq.~(\ref{result-janus-Tzero-2}), where $l$ plays the role of an infrared regulator.
At finite temperature, such expression generalizes to
\beq
\mathcal{C}^{F}_{\mathrm{Defect}}  =
\frac{4}{3} c \,  \eta(\gamma) \,
\log \le    \frac{1}{T \,  \delta} \ri  
 + \text{finite terms}   \, ,
 \label{eq:complexity_formation_defect}
\eeq
which is still logarithmically divergent.
However, in this case there is no need of an infrared regulator.
At finite temperature, it is also meaningful to consider the 
thermal complexity of formation of the Janus BTZ black hole in a geometry which already contains a defect.
In this case we compute the difference between the total complexity and its value at temperature $T=0,$ to obtain 
\beq
\mathcal{C}^{F}_{\mathrm{Thermal}}=
\frac{4}{3} c \,\eta(\gamma) \log \left( \frac{2}{\pi l T} \right) \, .
\eeq
Compared to Eq.~\eqref{eq:complexity_formation_defect}, this quantity is UV finite, but it requires  an IR regulator $l.$ 

We also numerically computed the time evolution of the volume complexity
for the time-dependent Janus BTZ black hole. In this case the boundary theory is not
an interface CFT,  but corresponds to two entangled CFTs with different
values of the dilaton field on each of the boundaries. At $t_B=0$, the boundary
theories start from an out-of-equilibrium state and the time-dependent Janus black hole
background is the gravity dual of the thermalisation process. 
The rate of growth of the volume as a function of the boundary time $t_B$ is shown
in figure \ref{fig_volume_BTZ_subtraction_gamma}. 

At late time, the growth rate of the volume saturates at the same constant (proportional to $T S$) for all the values of $\gamma$.
So the coupling does not influence the computational power of the CFT at equilibrium.
This could come as a surprise given that $\gamma$ determines
 the boundary values of the dilaton, which are dual to the couplings of the boundary CFTs.
We find that at early times, where
 the dual field theory is in an out-of-equilibrium state for $\gamma \neq 0$,
 the Janus deformation always
 decreases the complexity growth rate
compared to the BTZ case. 
 Being out of equilibrium decreases the computational power of the CFT.

In this paper we investigated several aspects of the CV conjecture
in AdS$_3$ Janus geometries. Several open questions call for further investigation:
\begin{itemize}
\item It would be interesting to extend our analysis of the complexity of the time-dependent Janus BTZ BH to more general geometries.
For regions made by the union of two segments on different sides of the Kruskal diagram,
 there is a phase transition in the topology of the HRT surface: 
 depending on the size of the segments, the HRT surface can be connected or disconnected \cite{Ugajin:2014nca}. 
This should be reflected in a discontinuity of subregion complexity 
as a function of the length of the segment; similar discontinuities 
in subregion complexity appear also for regions made by two segments
in AdS$_3$ Poincar\'e patch, see e.g. \cite{Abt:2017pmf,Auzzi:2019vyh}.
\item Another direction is the generalization to higher dimensions, for example to Janus AdS$_5$.
This is the topic of a follow-up work \cite{Baiguera:2021cba}. In this case the leading order divergence is 
power-like, and depends on the regularization method that is used. Subleading logarithmic divergences 
 instead turn out to be independent of the regularization method.
 This result points towards a universal property typical of the coefficient of the logarithmic divergences, as we also observed in the three-dimensional case considered here.
\item An important topic for further investigation is the complexity=action conjecture \cite{Brown:2015bva}.
For AdS$_3$ Randall-Sundrum \cite{Chapman:2018bqj} and BCFT  \cite{Sato:2019kik,Braccia:2019xxi}
the action complexity due to the defect is finite, and so it provides a situation where the CV and the CA
conjecture give radically different outcomes. In higher dimension BCFT, it turns out that CV and CA have instead 
the same leading-order divergence structure \cite{Sato:2019kik}. It would be interesting to investigate 
the case of the Janus interface, both in AdS$_3$ and  in AdS$_5$, to check if this behaviour is universal for defects.
\item It would be interesting to compare the holographic result with a direct calculation of complexity 
on the field theory side. In \cite{Braccia:2019xxi} circuit complexity was computed
in a free field theory with boundary, following the approach introduced in \cite{Jefferson:2017sdb}.
Even if the free field theory calculation and the holographic dual are not directly related,
it is interesting that the result is logarithmically divergent, as in the CV case.
The absence of logarithmic divergences in CA so seems to disfavour this proposal.
It would be interesting to perform the  calculation of the complexity of the defect 
also in conformal field theory, following the approach in \cite{Caputa:2018kdj,Erdmenger:2020sup,Chagnet:2021uvi}.
The CFT calculation should be more directly related to the holographic dual
and may give further insights on the problem of identifying the correct
holographic dual of quantum computational complexity.
\end{itemize}

\section*{Acknowledgments}

We thank G. Bruno De Luca for valuable discussions.
The authors S.B. acknowledge support from the Independent Research Fund Denmark 
grant number DFF-6108-00340 “Towards a deeper understanding of black holes
 with non-relativistic holography” and from DFF-FNU through grant number DFF-4002-00037.

\appendix
\section{Jacobi elliptic functions and elliptic integrals}
\label{elliptics}
We work with the standard Jacobi elliptic functions and elliptic integrals defined along the lines of \cite{Abram1964}. We use the incomplete elliptic integrals
\begin{align}
\elF{x}{m}&=\int_0^x\frac{d\th}{\sqrt{1-m\sin^2\th}}\;,  \noindent \\
\elE{x}{m}&=\int_0^x d\th\sqrt{1-m\sin^2\th}\;,\noindent \\
\Pi\left(n;x \left| m \right.\right) &=\int_0^x \frac{d\th}{\left(1-n\sin^2\th\right)\sqrt{1-m\sin^2\th}}.
\end{align}
of the first, second and third kind, respectively. The complete elliptic integrals are defined as
\beq
\elF{\frac{\pi}{2}}{m}=\eK(m)\;, \quad \elE{\frac{\pi}{2}}{m}=\eE(m)\;, \quad \Pi\left(n; \frac{\pi}{2}\Big| m \right)=\eP{n}{m}\;.
\eeq 
We also use the Jacobi amplitude $\vf=\text{am}(x|m)$ which is the inverse of $\elF{x}{m}$
\beq
x=\elF{\text{am}(x|m)}{m}\;.
\eeq
The Jacobi elliptic functions are defined as
\beq
\label{Jels}
\sn{x}{m}=\sin\vf,\quad\cn{x}{m}=\cos\vf\quad\text{and}\quad\dn{x}{m}=\sqrt{1-m\sin^2\vf},
\eeq
such that $\sn{\mathds{K}(m)}{m}=1$ and $\cn{\mathds{K}(m)}{m}=0$.

\bibliography{at}

\providecommand{\href}[2]{#2}\begin{thebibliography}{10}

\bibitem{Ryu:2006bv}
S.~Ryu and T.~Takayanagi, ``{Holographic derivation of entanglement entropy
  from AdS/CFT},'' {\em Phys. Rev. Lett.} {\bf 96} (2006) 181602,
  \href{http://arXiv.org/abs/hep-th/0603001}{{\tt hep-th/0603001}}.

\bibitem{Hubeny:2007xt}
V.~E. Hubeny, M.~Rangamani, and T.~Takayanagi, ``{A Covariant holographic
  entanglement entropy proposal},'' {\em JHEP} {\bf 07} (2007) 062,
  \href{http://arXiv.org/abs/0705.0016}{{\tt 0705.0016}}.

\bibitem{Lewkowycz:2013nqa}
A.~Lewkowycz and J.~Maldacena, ``{Generalized gravitational entropy},'' {\em
  JHEP} {\bf 08} (2013) 090, \href{http://arXiv.org/abs/1304.4926}{{\tt
  1304.4926}}.

\bibitem{Faulkner:2013ana}
T.~Faulkner, A.~Lewkowycz, and J.~Maldacena, ``{Quantum corrections to
  holographic entanglement entropy},'' {\em JHEP} {\bf 11} (2013) 074,
  \href{http://arXiv.org/abs/1307.2892}{{\tt 1307.2892}}.

\bibitem{Almheiri:2019psf}
A.~Almheiri, N.~Engelhardt, D.~Marolf, and H.~Maxfield, ``{The entropy of bulk
  quantum fields and the entanglement wedge of an evaporating black hole},''
  {\em JHEP} {\bf 12} (2019) 063, \href{http://arXiv.org/abs/1905.08762}{{\tt
  1905.08762}}.

\bibitem{Penington:2019npb}
G.~Penington, ``{Entanglement Wedge Reconstruction and the Information
  Paradox},'' {\em JHEP} {\bf 09} (2020) 002,
  \href{http://arXiv.org/abs/1905.08255}{{\tt 1905.08255}}.

\bibitem{Almheiri:2019hni}
A.~Almheiri, R.~Mahajan, J.~Maldacena, and Y.~Zhao, ``{The Page curve of
  Hawking radiation from semiclassical geometry},'' {\em JHEP} {\bf 03} (2020)
  149, \href{http://arXiv.org/abs/1908.10996}{{\tt 1908.10996}}.

\bibitem{Susskind:2014rva}
L.~Susskind, ``{Computational Complexity and Black Hole Horizons},'' {\em
  Fortsch. Phys.} {\bf 64} (2016) 24--43,
  \href{http://arXiv.org/abs/1403.5695}{{\tt 1403.5695}}. [Addendum:
  Fortsch.Phys. 64, 44--48 (2016)].

\bibitem{MIyaji:2015mia}
M.~Miyaji, T.~Numasawa, N.~Shiba, T.~Takayanagi, and K.~Watanabe, ``{Distance
  between Quantum States and Gauge-Gravity Duality},'' {\em Phys. Rev. Lett.}
  {\bf 115} (2015), no.~26, 261602, \href{http://arXiv.org/abs/1507.07555}{{\tt
  1507.07555}}.

\bibitem{Alishahiha:2015rta}
M.~Alishahiha, ``{Holographic Complexity},'' {\em Phys. Rev. D} {\bf 92}
  (2015), no.~12, 126009, \href{http://arXiv.org/abs/1509.06614}{{\tt
  1509.06614}}.

\bibitem{Stanford:2014jda}
D.~Stanford and L.~Susskind, ``{Complexity and Shock Wave Geometries},'' {\em
  Phys. Rev. D} {\bf 90} (2014), no.~12, 126007,
  \href{http://arXiv.org/abs/1406.2678}{{\tt 1406.2678}}.

\bibitem{Susskind:2014moa}
L.~Susskind, ``{Entanglement is not enough},'' {\em Fortsch. Phys.} {\bf 64}
  (2016) 49--71, \href{http://arXiv.org/abs/1411.0690}{{\tt 1411.0690}}.

\bibitem{Nielsen1}
M.~A. Nielsen, ``{A geometric approach to quantum circuit lower bounds},'' {\em
  Quantum Information and Computation} {\bf 6} (05, 2006)
  \href{http://arXiv.org/abs/quant-ph/0502070}{{\tt quant-ph/0502070}}.

\bibitem{Nielsen2}
M.~A.~N. Mark R.~Dowling, ``{The geometry of quantum computation},'' {\em
  Quantum Information and Computation} {\bf 8} (01, 2010) 861,
  \href{http://arXiv.org/abs/quant-ph/0701004}{{\tt quant-ph/0701004}}.

\bibitem{Brown:2016wib}
A.~R. Brown, L.~Susskind, and Y.~Zhao, ``{Quantum Complexity and Negative
  Curvature},'' {\em Phys. Rev. D} {\bf 95} (2017), no.~4, 045010,
  \href{http://arXiv.org/abs/1608.02612}{{\tt 1608.02612}}.

\bibitem{Brown:2019whu}
A.~R. Brown and L.~Susskind, ``{Complexity geometry of a single qubit},'' {\em
  Phys. Rev. D} {\bf 100} (2019), no.~4, 046020,
  \href{http://arXiv.org/abs/1903.12621}{{\tt 1903.12621}}.

\bibitem{Auzzi:2020idm}
R.~Auzzi, S.~Baiguera, G.~B. De~Luca, A.~Legramandi, G.~Nardelli, and
  N.~Zenoni, ``{Geometry of quantum complexity},'' {\em Phys. Rev. D} {\bf 103}
  (2021) 106021, \href{http://arXiv.org/abs/2011.07601}{{\tt 2011.07601}}.

\bibitem{Jefferson:2017sdb}
R.~Jefferson and R.~C. Myers, ``{Circuit complexity in quantum field theory},''
  {\em JHEP} {\bf 10} (2017) 107, \href{http://arXiv.org/abs/1707.08570}{{\tt
  1707.08570}}.

\bibitem{Chapman:2017rqy}
S.~Chapman, M.~P. Heller, H.~Marrochio, and F.~Pastawski, ``{Toward a
  Definition of Complexity for Quantum Field Theory States},'' {\em Phys. Rev.
  Lett.} {\bf 120} (2018), no.~12, 121602,
  \href{http://arXiv.org/abs/1707.08582}{{\tt 1707.08582}}.

\bibitem{Khan:2018rzm}
R.~Khan, C.~Krishnan, and S.~Sharma, ``{Circuit Complexity in Fermionic Field
  Theory},'' {\em Phys. Rev. D} {\bf 98} (2018), no.~12, 126001,
  \href{http://arXiv.org/abs/1801.07620}{{\tt 1801.07620}}.

\bibitem{Caputa:2018kdj}
P.~Caputa and J.~M. Magan, ``{Quantum Computation as Gravity},'' {\em Phys.
  Rev. Lett.} {\bf 122} (2019), no.~23, 231302,
  \href{http://arXiv.org/abs/1807.04422}{{\tt 1807.04422}}.

\bibitem{Erdmenger:2020sup}
J.~Erdmenger, M.~Gerbershagen, and A.-L. Weigel, ``{Complexity measures from
  geometric actions on Virasoro and Kac-Moody orbits},'' {\em JHEP} {\bf 11}
  (2020) 003, \href{http://arXiv.org/abs/2004.03619}{{\tt 2004.03619}}.

\bibitem{Chagnet:2021uvi}
N.~Chagnet, S.~Chapman, J.~de~Boer, and C.~Zukowski, ``{Complexity for
  Conformal Field Theories in General Dimensions},''
  \href{http://arXiv.org/abs/2103.06920}{{\tt 2103.06920}}.

\bibitem{Caputa:2017yrh}
P.~Caputa, N.~Kundu, M.~Miyaji, T.~Takayanagi, and K.~Watanabe, ``{Liouville
  Action as Path-Integral Complexity: From Continuous Tensor Networks to
  AdS/CFT},'' {\em JHEP} {\bf 11} (2017) 097,
  \href{http://arXiv.org/abs/1706.07056}{{\tt 1706.07056}}.

\bibitem{Parker:2018yvk}
D.~E. Parker, X.~Cao, A.~Avdoshkin, T.~Scaffidi, and E.~Altman, ``{A Universal
  Operator Growth Hypothesis},'' {\em Phys. Rev. X} {\bf 9} (2019), no.~4,
  041017, \href{http://arXiv.org/abs/1812.08657}{{\tt 1812.08657}}.

\bibitem{Barbon:2019wsy}
J.~L.~F. Barb\'on, E.~Rabinovici, R.~Shir, and R.~Sinha, ``{On The Evolution Of
  Operator Complexity Beyond Scrambling},'' {\em JHEP} {\bf 10} (2019) 264,
  \href{http://arXiv.org/abs/1907.05393}{{\tt 1907.05393}}.

\bibitem{Brown:2015bva}
A.~R. Brown, D.~A. Roberts, L.~Susskind, B.~Swingle, and Y.~Zhao,
  ``{Holographic Complexity Equals Bulk Action?},'' {\em Phys. Rev. Lett.} {\bf
  116} (2016), no.~19, 191301, \href{http://arXiv.org/abs/1509.07876}{{\tt
  1509.07876}}.

\bibitem{Brown:2015lvg}
A.~R. Brown, D.~A. Roberts, L.~Susskind, B.~Swingle, and Y.~Zhao,
  ``{Complexity, action, and black holes},'' {\em Phys. Rev. D} {\bf 93}
  (2016), no.~8, 086006, \href{http://arXiv.org/abs/1512.04993}{{\tt
  1512.04993}}.

\bibitem{Carmi:2017jqz}
D.~Carmi, S.~Chapman, H.~Marrochio, R.~C. Myers, and S.~Sugishita, ``{On the
  Time Dependence of Holographic Complexity},'' {\em JHEP} {\bf 11} (2017) 188,
  \href{http://arXiv.org/abs/1709.10184}{{\tt 1709.10184}}.

\bibitem{Carmi:2016wjl}
D.~Carmi, R.~C. Myers, and P.~Rath, ``{Comments on Holographic Complexity},''
  {\em JHEP} {\bf 03} (2017) 118, \href{http://arXiv.org/abs/1612.00433}{{\tt
  1612.00433}}.

\bibitem{Banados:1992gq}
M.~Banados, M.~Henneaux, C.~Teitelboim, and J.~Zanelli, ``{Geometry of the
  (2+1) black hole},'' {\em Phys. Rev. D} {\bf 48} (1993) 1506--1525,
  \href{http://arXiv.org/abs/gr-qc/9302012}{{\tt gr-qc/9302012}}. [Erratum:
  Phys.Rev.D 88, 069902 (2013)].

\bibitem{Abt:2017pmf}
R.~Abt, J.~Erdmenger, H.~Hinrichsen, C.~M. Melby-Thompson, R.~Meyer, C.~Northe,
  and I.~A. Reyes, ``{Topological Complexity in AdS$_3$/CFT$_2$},'' {\em
  Fortsch. Phys.} {\bf 66} (2018), no.~6, 1800034,
  \href{http://arXiv.org/abs/1710.01327}{{\tt 1710.01327}}.

\bibitem{Auzzi:2019vyh}
R.~Auzzi, S.~Baiguera, A.~Legramandi, G.~Nardelli, P.~Roy, and N.~Zenoni, ``{On
  subregion action complexity in AdS$_{3}$ and in the BTZ black hole},'' {\em
  JHEP} {\bf 01} (2020) 066, \href{http://arXiv.org/abs/1910.00526}{{\tt
  1910.00526}}.

\bibitem{Abt:2018ywl}
R.~Abt, J.~Erdmenger, M.~Gerbershagen, C.~M. Melby-Thompson, and C.~Northe,
  ``{Holographic Subregion Complexity from Kinematic Space},'' {\em JHEP} {\bf
  01} (2019) 012, \href{http://arXiv.org/abs/1805.10298}{{\tt 1805.10298}}.

\bibitem{Agon:2018zso}
C.~A. Ag\'on, M.~Headrick, and B.~Swingle, ``{Subsystem Complexity and
  Holography},'' {\em JHEP} {\bf 02} (2019) 145,
  \href{http://arXiv.org/abs/1804.01561}{{\tt 1804.01561}}.

\bibitem{Auzzi:2019fnp}
R.~Auzzi, S.~Baiguera, A.~Mitra, G.~Nardelli, and N.~Zenoni, ``{Subsystem
  complexity in warped AdS},'' {\em JHEP} {\bf 09} (2019) 114,
  \href{http://arXiv.org/abs/1906.09345}{{\tt 1906.09345}}.

\bibitem{Chen:2018mcc}
B.~Chen, W.-M. Li, R.-Q. Yang, C.-Y. Zhang, and S.-J. Zhang, ``{Holographic
  subregion complexity under a thermal quench},'' {\em JHEP} {\bf 07} (2018)
  034, \href{http://arXiv.org/abs/1803.06680}{{\tt 1803.06680}}.

\bibitem{Auzzi:2019mah}
R.~Auzzi, G.~Nardelli, F.~I. Schaposnik~Massolo, G.~Tallarita, and N.~Zenoni,
  ``{On volume subregion complexity in Vaidya spacetime},'' {\em JHEP} {\bf 11}
  (2019) 098, \href{http://arXiv.org/abs/1908.10832}{{\tt 1908.10832}}.

\bibitem{Caceres:2019pgf}
E.~Caceres, S.~Chapman, J.~D. Couch, J.~P. Hernandez, R.~C. Myers, and S.-M.
  Ruan, ``{Complexity of Mixed States in QFT and Holography},'' {\em JHEP} {\bf
  03} (2020) 012, \href{http://arXiv.org/abs/1909.10557}{{\tt 1909.10557}}.

\bibitem{Calabrese:2004eu}
P.~Calabrese and J.~L. Cardy, ``{Entanglement entropy and quantum field
  theory},'' {\em J. Stat. Mech.} {\bf 0406} (2004) P06002,
  \href{http://arXiv.org/abs/hep-th/0405152}{{\tt hep-th/0405152}}.

\bibitem{Affleck:1991tk}
I.~Affleck and A.~W.~W. Ludwig, ``{Universal noninteger 'ground state
  degeneracy' in critical quantum systems},'' {\em Phys. Rev. Lett.} {\bf 67}
  (1991) 161--164.

\bibitem{Randall:1999vf}
L.~Randall and R.~Sundrum, ``{An Alternative to compactification},'' {\em Phys.
  Rev. Lett.} {\bf 83} (1999) 4690--4693,
  \href{http://arXiv.org/abs/hep-th/9906064}{{\tt hep-th/9906064}}.

\bibitem{Azeyanagi:2007qj}
T.~Azeyanagi, A.~Karch, T.~Takayanagi, and E.~G. Thompson, ``{Holographic
  calculation of boundary entropy},'' {\em JHEP} {\bf 03} (2008) 054,
  \href{http://arXiv.org/abs/0712.1850}{{\tt 0712.1850}}.

\bibitem{Bhattacharya:2021jrn}
A.~Bhattacharya, A.~Bhattacharyya, P.~Nandy, and A.~K. Patra, ``{Islands and
  complexity of eternal black hole and radiation subsystems for a doubly
  holographic model},'' {\em JHEP} {\bf 05} (2021) 135,
  \href{http://arXiv.org/abs/2103.15852}{{\tt 2103.15852}}.

\bibitem{Takayanagi:2011zk}
T.~Takayanagi, ``{Holographic Dual of BCFT},'' {\em Phys. Rev. Lett.} {\bf 107}
  (2011) 101602, \href{http://arXiv.org/abs/1105.5165}{{\tt 1105.5165}}.

\bibitem{Fujita:2011fp}
M.~Fujita, T.~Takayanagi, and E.~Tonni, ``{Aspects of AdS/BCFT},'' {\em JHEP}
  {\bf 11} (2011) 043, \href{http://arXiv.org/abs/1108.5152}{{\tt 1108.5152}}.

\bibitem{Flory:2017ftd}
M.~Flory, ``{A complexity/fidelity susceptibility $g$-theorem for
  AdS$_{3}$/BCFT$_{2}$},'' {\em JHEP} {\bf 06} (2017) 131,
  \href{http://arXiv.org/abs/1702.06386}{{\tt 1702.06386}}.

\bibitem{Bak:2003jk}
D.~Bak, M.~Gutperle, and S.~Hirano, ``{A Dilatonic deformation of AdS(5) and
  its field theory dual},'' {\em JHEP} {\bf 05} (2003) 072,
  \href{http://arXiv.org/abs/hep-th/0304129}{{\tt hep-th/0304129}}.

\bibitem{Bak:2007jm}
D.~Bak, M.~Gutperle, and S.~Hirano, ``{Three dimensional Janus and
  time-dependent black holes},'' {\em JHEP} {\bf 02} (2007) 068,
  \href{http://arXiv.org/abs/hep-th/0701108}{{\tt hep-th/0701108}}.

\bibitem{Nakata:2020fjg}
Y.~Nakata, T.~Takayanagi, Y.~Taki, K.~Tamaoka, and Z.~Wei, ``{Holographic
  Pseudo Entropy},'' \href{http://arXiv.org/abs/2005.13801}{{\tt 2005.13801}}.

\bibitem{Chapman:2018bqj}
S.~Chapman, D.~Ge, and G.~Policastro, ``{Holographic Complexity for Defects
  Distinguishes Action from Volume},'' {\em JHEP} {\bf 05} (2019) 049,
  \href{http://arXiv.org/abs/1811.12549}{{\tt 1811.12549}}.

\bibitem{Sato:2019kik}
Y.~Sato and K.~Watanabe, ``{Does Boundary Distinguish Complexities?},'' {\em
  JHEP} {\bf 11} (2019) 132, \href{http://arXiv.org/abs/1908.11094}{{\tt
  1908.11094}}.

\bibitem{Braccia:2019xxi}
P.~Braccia, A.~L. Cotrone, and E.~Tonni, ``{Complexity in the presence of a
  boundary},'' {\em JHEP} {\bf 02} (2020) 051,
  \href{http://arXiv.org/abs/1910.03489}{{\tt 1910.03489}}.

\bibitem{Bak:2015jxd}
D.~Bak, ``{Information metric and Euclidean Janus correspondence},'' {\em Phys.
  Lett. B} {\bf 756} (2016) 200--204,
  \href{http://arXiv.org/abs/1512.04735}{{\tt 1512.04735}}.

\bibitem{Mazhari:2016yng}
N.~Mazhari, D.~Momeni, S.~Bahamonde, M.~Faizal, and R.~Myrzakulov,
  ``{Holographic Complexity and Fidelity Susceptibility as Holographic
  Information Dual to Different Volumes in AdS},'' {\em Phys. Lett. B} {\bf
  766} (2017) 94--101, \href{http://arXiv.org/abs/1609.00250}{{\tt
  1609.00250}}.

\bibitem{Freedman:2003ax}
D.~Z. Freedman, C.~Nunez, M.~Schnabl, and K.~Skenderis, ``{Fake supergravity
  and domain wall stability},'' {\em Phys. Rev. D} {\bf 69} (2004) 104027,
  \href{http://arXiv.org/abs/hep-th/0312055}{{\tt hep-th/0312055}}.

\bibitem{Papadimitriou:2004rz}
I.~Papadimitriou and K.~Skenderis, ``{Correlation functions in holographic RG
  flows},'' {\em JHEP} {\bf 10} (2004) 075,
  \href{http://arXiv.org/abs/hep-th/0407071}{{\tt hep-th/0407071}}.

\bibitem{Chiodaroli:2010ur}
M.~Chiodaroli, M.~Gutperle, and L.-Y. Hung, ``{Boundary entropy of
  supersymmetric Janus solutions},'' {\em JHEP} {\bf 09} (2010) 082,
  \href{http://arXiv.org/abs/1005.4433}{{\tt 1005.4433}}.

\bibitem{Bak:2011ga}
D.~Bak, M.~Gutperle, and R.~A. Janik, ``{Janus Black Holes},'' {\em JHEP} {\bf
  10} (2011) 056, \href{http://arXiv.org/abs/1109.2736}{{\tt 1109.2736}}.

\bibitem{Bak:2020enw}
D.~Bak, C.~Kim, S.-H. Yi, and J.~Yoon, ``{Unitarity of Entanglement and Islands
  in Two-Sided Janus Black Holes},'' {\em JHEP} {\bf 01} (2021) 155,
  \href{http://arXiv.org/abs/2006.11717}{{\tt 2006.11717}}.

\bibitem{Estes:2014hka}
J.~Estes, K.~Jensen, A.~O'Bannon, E.~Tsatis, and T.~Wrase, ``{On Holographic
  Defect Entropy},'' {\em JHEP} {\bf 05} (2014) 084,
  \href{http://arXiv.org/abs/1403.6475}{{\tt 1403.6475}}.

\bibitem{Bak:2016rpn}
D.~Bak, A.~Gustavsson, and S.-J. Rey, ``{Conformal Janus on Euclidean
  Sphere},'' {\em JHEP} {\bf 12} (2016) 025,
  \href{http://arXiv.org/abs/1605.00857}{{\tt 1605.00857}}.

\bibitem{Gutperle:2016gfe}
M.~Gutperle and A.~Trivella, ``{Note on entanglement entropy and regularization
  in holographic interface theories},'' {\em Phys. Rev. D} {\bf 95} (2017),
  no.~6, 066009, \href{http://arXiv.org/abs/1611.07595}{{\tt 1611.07595}}.

\bibitem{Chapman:2016hwi}
S.~Chapman, H.~Marrochio, and R.~C. Myers, ``{Complexity of Formation in
  Holography},'' {\em JHEP} {\bf 01} (2017) 062,
  \href{http://arXiv.org/abs/1610.08063}{{\tt 1610.08063}}.

\bibitem{Bak:2007qw}
D.~Bak, M.~Gutperle, and A.~Karch, ``{Time dependent black holes and thermal
  equilibration},'' {\em JHEP} {\bf 12} (2007) 034,
  \href{http://arXiv.org/abs/0708.3691}{{\tt 0708.3691}}.

\bibitem{Ugajin:2014nca}
Y.~Nakaguchi, N.~Ogawa, and T.~Ugajin, ``{Holographic Entanglement and Causal
  Shadow in Time-Dependent Janus Black Hole},'' {\em JHEP} {\bf 07} (2015) 080,
  \href{http://arXiv.org/abs/1412.8600}{{\tt 1412.8600}}.

\bibitem{Headrick:2014cta}
M.~Headrick, V.~E. Hubeny, A.~Lawrence, and M.~Rangamani, ``{Causality
  \textbackslash{}\& holographic entanglement entropy},'' {\em JHEP} {\bf 12}
  (2014) 162, \href{http://arXiv.org/abs/1408.6300}{{\tt 1408.6300}}.

\bibitem{Baiguera:2021cba}
S.~Baiguera, S.~Bonansea, and K.~Toccacelo, ``{Volume complexity for the
  non-supersymmetric Janus AdS$_5$ geometry},''
  \href{http://arXiv.org/abs/2105.12743}{{\tt 2105.12743}}.

\bibitem{Abram1964}
M.~Abramowitz and I.~Stegun, {\em Handbook of Mathematical Functions}.
\newblock fifth edition, Dover (1964), New York.

\end{thebibliography}
\bibliographystyle{at}

\end{document}